\documentclass[11pt,twocolumn]{article}

\usepackage{graphicx}
\usepackage{algorithm,algorithmic}
\usepackage{amsmath}
\usepackage{multirow}
\usepackage{float}
\usepackage{url}
\usepackage[none]{hyphenat}
\usepackage{authblk}

\usepackage{txfonts,balance}

\usepackage[hyphenbreaks]{breakurl}

\usepackage{afterpage}
\newcommand\T{\rule{0pt}{2.6ex}}

\begin{document}
\newcounter{cntr1}
\newcounter{cntr2}
\emergencystretch 3em

\title{Circuit simulation using explicit methods}

\author[1]{Mahesh~B.~Patil}
\author[2]{V.V.S.~Pavan~Kumar~Hari}
\affil[1]{Department of Electrical Engineering, Indian Institute of Technology Bombay}
\affil[2]{Department of Energy Science and Engineering, Indian Institute of Technology Bombay}

\maketitle

\begin{abstract}
Use of explicit methods for simulating electrical circuits,
especially for power electronics applications, is described.
Application of the forward Euler method to a half-wave
rectifier is discussed, and the limitations of a fixed-step
method are pointed out. Implementation of the Runge-Kutta-Fehlberg (RKF) method,
which allows variable time steps, for the half-wave rectifier circuit
is discussed, and its advantages pointed out. Formulation of circuit equations
for the purpose of simulation using the RKF method is described
for some more examples. Stability and accuracy issues related to power
electronic circuits are brought out, and mechanisms to address them
are presented. Future plans related to this work are described.
\end{abstract}

\section{Introduction}
Circuit simulation packages (e.g.,
NGSPICE\,\cite{ngspice},
PSIM\,\cite{psim},
PSCAD\,\cite{pscad})
generally employ implicit methods, such as backward Euler
and trapezoidal, for discretising the circuit equations involving
time derivatives. Since these implicit methods are unconditionally stable,
the simulator time steps are not constrained by stability
issues. In other words, with these methods, the time step size
is determined by accuracy considerations rather than stability
constraints\,\cite{mbpvrvtr}.
Therefore, even though they involve more work per time step,
implicit methods are still the only practical option for circuits
with very small time constants.

On the other hand, explicit methods, such as the Runge-Kutta-Fehlberg
(RKF)\,\cite{burden}
and Dormand-Prince (DP)\,\cite{dormand} 4-5 pairs, involve less work per time step, and
are ideally suited for non-stiff systems in which the time constants are not too small.
Unfortunately, many circuits of practical interest do involve small time constants
arising from the small on-state resistance of a switch or a small capacitance
associated with a semiconductor device. In such cases, a simulator using an
explicit method is forced to take very small time steps although accuracy
considerations would allow much larger time steps. Application of explicit
methods to electronic circuits is therefore severely limited.

Power electronic circuits offer a unique opportunity to utilise the
efficiency~-- due to smaller computational effort per time step~--
of explicit methods if the switches are treated as ideal. The simulator
PLECS\,\cite{plecs} exploits this idea to enable fast simulation of a
variety of power electronic circuits. However, owing probably to the
commercial nature of PLECS, hardly any details are available in the
public domain about how the ideas broadly described in the PLECS\,4.6
manual are actually implemented.

It is the purpose of this paper to discuss various issues related
to the use of explicit methods for power electronic circuit
simulation. We consider a few examples to illustrate the basic
implementation scheme, the various challenges, and some ways to
address them.
\section{Half-wave rectifier}
\label{sec_hwrect}
\begin{figure}[!ht]
\centering
\scalebox{0.9}{\includegraphics{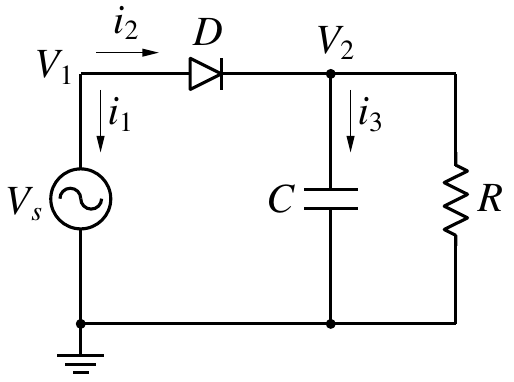}}
\vspace*{-0.2cm}
\caption{Half-wave rectifier circuit. The source voltage is
given by $V_s \,$=$\, V_m\,\sin \omega t$.}
\label{fig_hwr_1}
\end{figure}

A half-wave rectifier circuit is shown in Fig.~\ref{fig_hwr_1}.
In the following, we discuss how this circuit can be handled
using an implicit method (backward Euler), an explicit method
with a fixed time step (forward Euler), and an explicit method
with a variable time step (RKF). The parameter values considered
in all cases are:
$V_m \,$=$\, 10$\,V,
$f \,$=$\, 50$\,Hz, $C \,$=$\, 1$\,mF, $R \,$=$\, 200\,\Omega$.
The on-state voltage drop for the diode is taken to be
$V_{\mathrm{on}} \,$=$\, 0$\,V.

\subsection{Backward Euler (BE) method}
\label{sec_hwrect_be}
The capacitor equation,
$\displaystyle\frac{dV_C}{dt} \,$=$\, \displaystyle\frac{i_C}{C}$
is discretised using the BE method as,
\begin{equation}
\displaystyle\frac{V_{C,n+1}-V_{C,n}}{h} = \displaystyle\frac{i_{C,n+1}}{C},
\label{eq_hwrect_be_1}
\end{equation}
where
$X_{n}$, $X_{n+1}$
denote the numerically obtained value of $X$ at
$t \,$=$\, t_n$, $t \,$=$\, t_{n+1}$, respectively. In the context of the half-wave
rectifier circuit, Eq.~\ref{eq_hwrect_be_1} gives,
\begin{equation}
i_{C,n+1} = \displaystyle\frac{C}{h}\,\left(V_{2,n+1}-V_{2,n}\right).
\label{eq_hwrect_be_2}
\end{equation}
Using Eq.~\ref{eq_hwrect_be_2} and the Modified Nodal Analysis
(MNA) approach\,\cite{mccalla} for assembling circuit equations,
we get
\begin{equation}
i_{1,n+1} + G_D\,(V_{1,n+1} - V_{2,n+1}) = 0,
\label{eq_hwrect_be_3}
\end{equation}
\begin{equation}
G_D\,(V_{2,n+1} - V_{1,n+1}) + \displaystyle\frac{C}{h}\,\left[V_{2,n+1}-V_{2,n}\right] + G\,V_{2,n+1} = 0,
\label{eq_hwrect_be_4}
\end{equation}
\begin{equation}
V_{1,n+1} - V_m\sin \omega t_{n+1} = 0,
\label{eq_hwrect_be_5}
\end{equation}
where $G \,$=$\, 1/R$. The diode has been modelled as $G_D \,$=$\, 1/R_D$,
where
$R_D \,$=$\, R_{\mathrm{on}}$ if the diode is conducting, and
$R_D \,$=$\, R_{\mathrm{off}}$ if it is not. By making
$R_{\mathrm{on}}$ and
$R_{\mathrm{off}}$
sufficiently small and large, respectively, ideal diode behaviour can
be realised.

Note that
Eqs.~\ref{eq_hwrect_be_3}-\ref{eq_hwrect_be_5}
form a nonlinear set of equations (in
$V_{1,n+1}$,
$V_{2,n+1}$,
$i_{1,n+1}$)
since $G_D$ is a nonlinear function of the diode voltage
$(V_1-V_2)$. To solve this system of equations, circuit
simulators generally employ the Newton-Raphson (N-R) iterative
method in which the Jacobian equation must be solved for a few
N-R iterations at each time point, thus making the procedure
computationally expensive.

On the positive side, the backward Euler method is unconditionally
stable and therefore does not impose an upper limit on the time step
from the stability perspective. Furthermore, it allows complicated
semiconductor device models to be incorporated in a straightforward
manner.

\subsection{Forward Euler (FE) method}
\label{sec_hwrect_fe}
The capacitor equation for the circuit of Fig.~\ref{fig_hwr_1} can be
discretised using the FE method as,
\begin{equation}
\displaystyle\frac{V_{2,n+1} - V_{2,n}}{h} = \displaystyle\frac{i_{3,n}}{C}.
\label{eq_hwrect_fe_1}
\end{equation}
Since
$V_{2,n}$,
$i_{3,n}$
are already known,
$V_{2,n+1}$
can be found simply by evaluation, i.e.,
\begin{equation}
V_{2,n+1} = V_{2,n} + \displaystyle\frac{h}{C}\,i_{3,n}.
\label{eq_hwrect_fe_2}
\end{equation}
We can now obtain the rest of the variables at $t_{n+1}$ by writing the
circuit equations, treating
$V_{2,n+1}$
as a {\it{known}} entity.
The resulting equations can be written, for example, as
\begin{equation}
V_{1,n+1} - V_m \sin \omega \,t_{n+1} = 0,
\label{eq_hwrect_fe_3a}
\end{equation}
\begin{equation}
i_{2,n+1} - G_D\,\left[V_{1,n+1} - V_{2,n+1}\right] = 0,
\label{eq_hwrect_fe_3b}
\end{equation}
\begin{equation}
i_{2,n+1} - i_{3,n+1} - G\,V_{2,n+1} = 0,
\label{eq_hwrect_fe_3c}
\end{equation}
a system of three equations in three variables:
$V_{1,n+1}$,
$i_{2,n+1}$,
$i_{3,n+1}$.
Note that $G_D$ is still a nonlinear function of
$(V_{1,n+1} - V_{2,n+1})$,
making the situation similar to the
backward Euler case.

\begin{figure}[!ht]
\centering
\scalebox{0.9}{\includegraphics{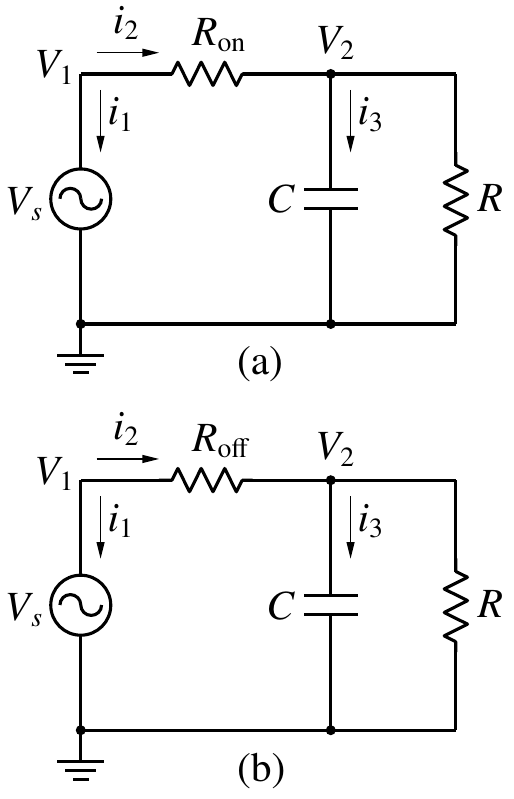}}
\vspace*{-0.2cm}
\caption{Circuit of Fig.~\ref{fig_hwr_1} in two situations:
(a)\,diode on, (b)\, diode off.}
\label{fig_hwr_2}
\end{figure}

In the interest of reducing the computation effort per time step,
we now replace the original circuit with two circuits, one with
the diode on, the other with the diode off,
as shown in Fig.~\ref{fig_hwr_2}.
Each of the two circuits is linear, with an associated set of
equations
(Eqs.~\ref{eq_hwrect_fe_3a}-\ref{eq_hwrect_fe_3c}), with
$R_D \,$=$\, R_{\mathrm{on}}$ in one case, and
$R_D \,$=$\, R_{\mathrm{off}}$ in the other.
Eqs.~\ref{eq_hwrect_fe_3a}-\ref{eq_hwrect_fe_3c} can be written
in the form
${\bf{A}}\,{\bf{x}} \,$=$\, {\bf{b}}$.
For example, with the diode on, we have
\begin{equation}
 \left[
 \begin{array}{ccc}
     1 & 0 & 0 \cr
     -G_D^{\mathrm{on}} & 1 & 0 \cr
     0 & 1 & -1 \cr
 \end{array}
 \right]
 \thinspace
 \left[
 \begin{array}{c}
   V_{1,n+1} \cr
   i_{2,n+1} \cr
   i_{3,n+1} \cr
 \end{array}
 \right]
 \thinspace =
 \left[
 \begin{array}{c}
   V_m \sin \omega \,t_{n+1} \cr
   -G_D^{\mathrm{on}}\,V_{2,n+1} \cr
   G\,V_{2,n+1} \cr
 \end{array}
 \right]~,
\label{eq_hwrect_fe_4}
\end{equation}
where
$G_D^{\mathrm{on}} \,$=$\, 1/R_{\mathrm{on}}$.
Note that the time step $h$ is not involved in the equations,
and therefore the ${\bf{A}}$ matrices are independent of time.
This has a major implication, viz.,
${\bf{A}}$ and
${\bf{A}}^{-1}$
for a specific switch configuration needs to be computed only once,
offering a very significant speed-up with respect to an implicit
method.

In some circuits, replacing a switch with
$R_{\mathrm{on}}$ or
$R_{\mathrm{off}}$
can create stability issues because
$R_{\mathrm{on}}$, a small resistance, can result in small $RC$-type
time constants, and
$R_{\mathrm{off}}$, a large resistance, in small $L/R$-type time
constants. In such cases, the simulator would be forced to take
very small time steps. It is therefore preferable to make
$R_{\mathrm{on}} \,$=$\, 0$ and
$R_{\mathrm{off}} \,$=$\, \infty$, i.e., replace the switch with a
short or open circuit.

\begin{figure*}[!ht]
\centering
\scalebox{0.9}{\includegraphics{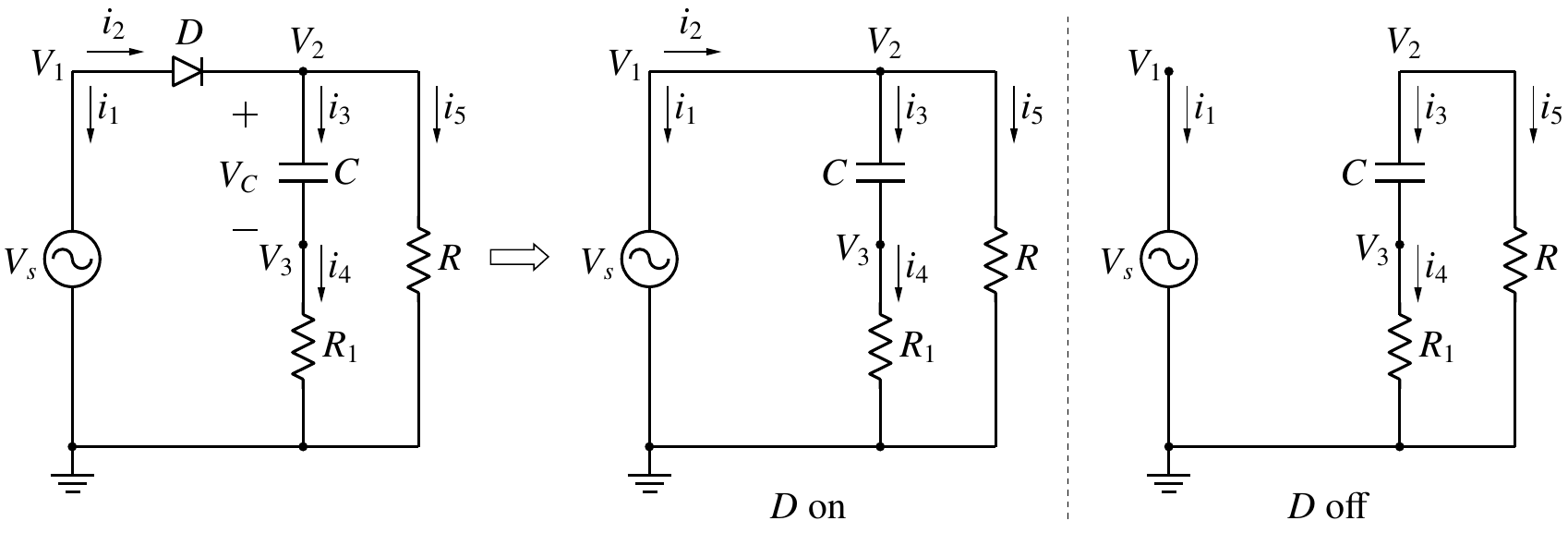}}
\vspace*{-0.2cm}
\caption{Addition of $R_1$ in series with $C$ in the half-wave rectifier
circuit of Fig.~\ref{fig_hwr_1}, and the equivalent circuits when the diode
is on or off.}
\label{fig_hwr_3}
\end{figure*}

In the half-wave rectifier circuit, using
$R_{\mathrm{on}} \,$=$\, 0$ creates a problem: The voltage source
comes directly across the capacitor, and the ${\bf{A}}$ matrix
becomes singular. A slight modification of the circuit, viz., addition
of a resistance $R_1$ in series with the capacitor, as shown in Fig.~\ref{fig_hwr_3},
circumvents this problem. The resistance $R_1$ should be small (taken as
10\,m$\Omega$ in this paper) so that it does not change the circuit operation.

There are different ways in which the set of equations for the on and off
circuits of Fig.~\ref{fig_hwr_3} can be written. In the following, we opt
for a systematic procedure which can be easily extended to other circuits.
We write the KCL equations at all nodes except the reference node,
followed by the element equations. Note that
\begin{equation}
V_{C,n+1} = V_{C,n} + \displaystyle\frac{1}{C}\,i_{3,n}
\label{eq_hwrect_fe_5}
\end{equation}
is treated as a known value. The set of equations is then given by
\begin{equation}
\label{eq_hwrect_fe_6_1}
i_{1,n+1} + i_{2,n+1} = 0,
\end{equation}
\begin{equation}
-i_{2,n+1} + i_{3,n+1} + i_{5,n+1} = 0,
\label{eq_hwrect_fe_6_2}
\end{equation}
\begin{equation}
-i_{3,n+1} + i_{4,n+1} = 0,
\label{eq_hwrect_fe_6_3}
\end{equation}
\begin{equation}
V_{1,n+1} = V_m \sin \omega \,t_{n+1},
\label{eq_hwrect_fe_6_4}
\end{equation}
\begin{equation}
V_{2,n+1} - V_{3,n+1} = V_{C,n+1},
\label{eq_hwrect_fe_6_5}
\end{equation}
\begin{equation}
i_{4,n+1} - G_1 V_{3,n+1} = 0,
\label{eq_hwrect_fe_6_6}
\end{equation}
\begin{equation}
i_{5,n+1} - G V_{2,n+1} = 0,
\label{eq_hwrect_fe_6_7}
\end{equation}
\begin{equation}
\begin{array}{cl}
V_{1,n+1} - V_{2,n+1} = 0 &{\textrm{if}}~D~{\textrm{is on}}, \\
i_{2,n+1} = 0 &{\textrm{if}}~D~{\textrm{is off}}.
\end{array}
\label{eq_hwrect_fe_6_8}
\end{equation}

Eqs.~\ref{eq_hwrect_fe_6_1}-\ref{eq_hwrect_fe_6_8} 
can be described in a matrix form,
${\bf{A}}\,{\bf{x}} \,$=$\, {\bf{b}}$,
where the solution vector {\bf{x}} consists of
$V_{1,n+1}$,
$V_{2,n+1}$,
$V_{3,n+1}$,
$i_{1,n+1}$,
$i_{2,n+1}$,
$i_{3,n+1}$,
$i_{4,n+1}$,
$i_{5,n+1}$,
and {\bf{b}} includes $V_{C,n+1}$ and $V_{s,n+1}$, both being known
values. Since the circuit has one switch, we have two possible {\bf{A}}
matrices,
${\bf{A}}^{\mathrm{(on)}}$ and
${\bf{A}}^{\mathrm{(off)}}$,
corresponding to the diode in the on and off state, respectively.
In general, the number of {\bf{A}} matrices would be $2^N$,
where $N$ is the number of switches in the circuit. After solving
${\bf{A}}\,{\bf{x}} \,$=$\, {\bf{b}}$,
we need to make sure that the diode state is consistent with the {\bf{A}}
matrix which was used in the computation. For example, in the half-wave
rectifier case, if
${\bf{A}}^{\mathrm{(off)}}$
was used, then $i_{2,n+1}$ must turn out to be positive. If it is not, we
need to reject the solution, use
${\bf{A}}^{\mathrm{(on)}}$
instead, and solve
${\bf{A}}\,{\bf{x}} \,$=$\, {\bf{b}}$
again. If
${\bf{A}}^{\mathrm{(on)}}$
has already been computed (and stored) earlier, solving
${\bf{A}}\,{\bf{x}} \,$=$\, {\bf{b}}$
takes a small computational effort.

Note that it is not necessary to pre-compute the
{\bf{A}} and ${\bf{A}}^{-1}$
matrices for all $2^N$ switch configurations. In may power electronic
circuits, only a small fraction of the 
switch configurations get actually visited, and pre-computing (and storing)
all possible
{\bf{A}} and ${\bf{A}}^{-1}$
matrices would be wasteful in terms of time and memory. Instead, the matrices
should be computed and stored dynamically, i.e., whenever a specific switch
configuration is encountered. The overall procedure can be summarised as shown
in Algorithm 1. We will refer to this scheme of using an explicit method for
electrical circuits as the ELEX (ELectrical EXplicit) scheme. For reasons
explained in the following, the FE method is not a good candidate for the
ELEX scheme. Instead, a variable time-step method (for example, RKF or DP)
would be preferred.
\begin{algorithm}
 \caption{ELEX scheme}
 \begin{algorithmic}[1]
  \STATE Initialise switch state vector ({\bf{S}}).
  \STATE $t \gets 0$
  \WHILE{$t < t_{\mathrm{end}}$}
    \STATE Update source outputs.
    \STATE Update state variables.
    \STATE Compute ${\bf{b}}$
    \IF{$\nexists {\bf{A}}({\bf{S}})$} \label{marker}
      \STATE Compute ${\bf{A}}({\bf{S}})$ and ${\bf{A}}^{-1}({\bf{S}})$.
    \ENDIF
    \STATE Solve ${\bf{A}} {\bf{x}} = {\bf{b}} $.
    \STATE Compute ${\bf{S}}_{\mathrm{new}}$ (new switch states)
    \IF{${\bf{S}} \neq {\bf{S}}_{\mathrm{new}}$}
      \STATE Pick new ${\bf{S}}$.
      \STATE Go to \ref{marker}
    \ENDIF
  \ENDWHILE
 \end{algorithmic}
\end{algorithm}

We now comment on stability issues related to the FE method when applied to
the circuit of Fig.~\ref{fig_hwr_3}.
The charging and discharging time constants, $\tau _1$ and $\tau _2$,
are given by
\begin{equation}
\tau _1 = (R_1\parallel R)\,C \simeq R_1C
= 10\,\mu{\textrm{sec}},
\end{equation}
\begin{equation}
\tau _2 = (R_1 + R)\,C \simeq R\,C
= 0.2\,{\textrm{sec}}.
\end{equation}

The FE method becomes unstable when the time step $h$ is larger than $2\,\tau$. Clearly, we need
to use $h$ smaller than $2\,\tau _1$ or 20\,$\mu$sec in the charging phase. Since the FE
method is a fixed time step method, we end up using this small time step (20\,$\mu$sec)
{\it{throughout}} the simulation, requiring
at least $T/\tau _1 \,$=$\, 20\,{\textrm{msec}}/20\,\mu{\textrm{sec}}$ or $10^3$ time points
in one period. From the accuracy point, such a large number of time steps is not required,
and the FE method is therefore too slow for many applications.

\begin{figure}[!ht]
\centering
{\includegraphics[width=0.49\textwidth]{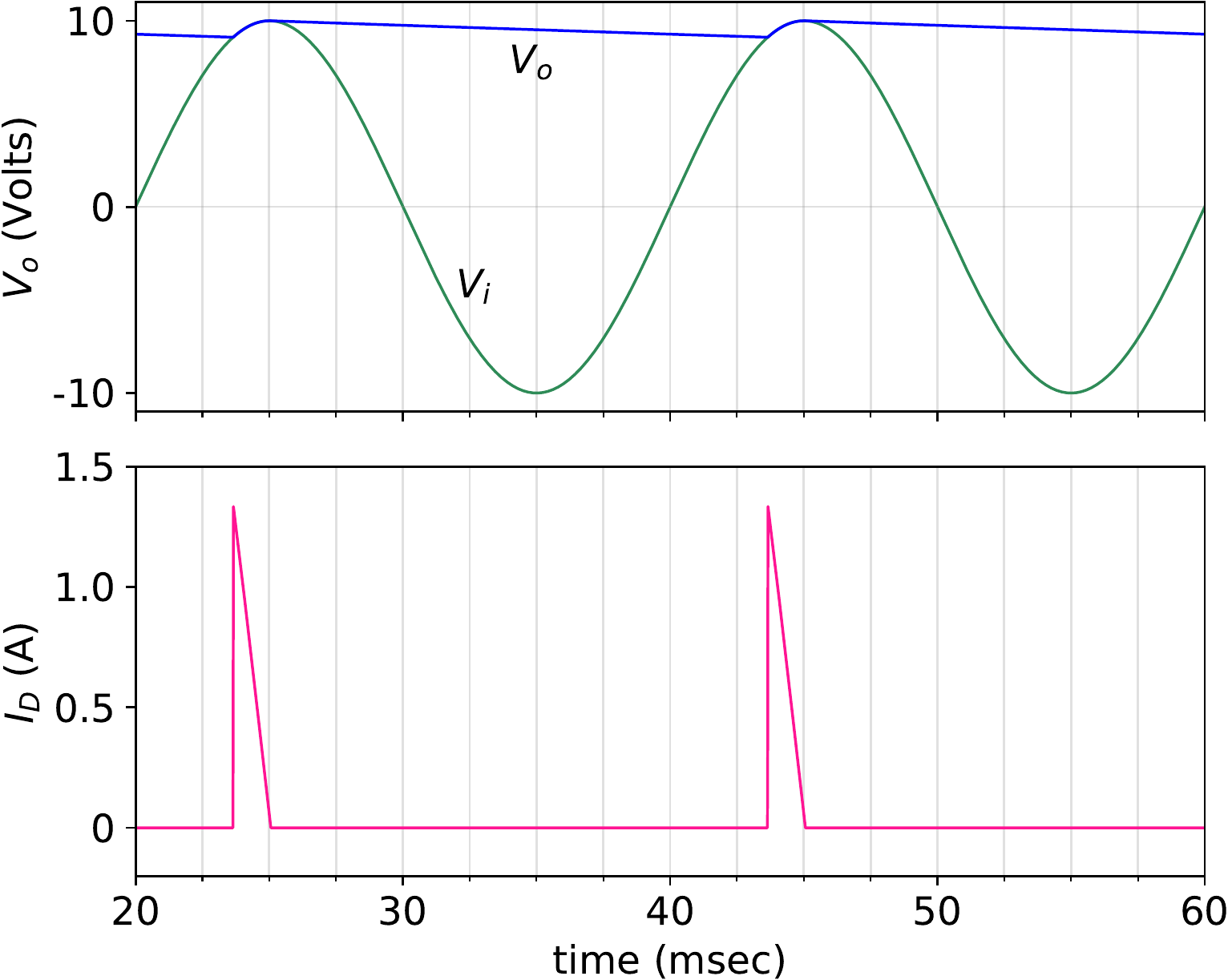}}
\vspace*{-0.7cm}
\caption{Input and output voltages, and diode current obtained with the FE
method for the half-wave rectifier circuit of Fig.~\ref{fig_hwr_3} with time
step $h \,$=$\, 10\,\mu$sec.}
\label{fig_hwr_4}
\end{figure}

\begin{figure}[!ht]
\centering
{\includegraphics[width=0.49\textwidth]{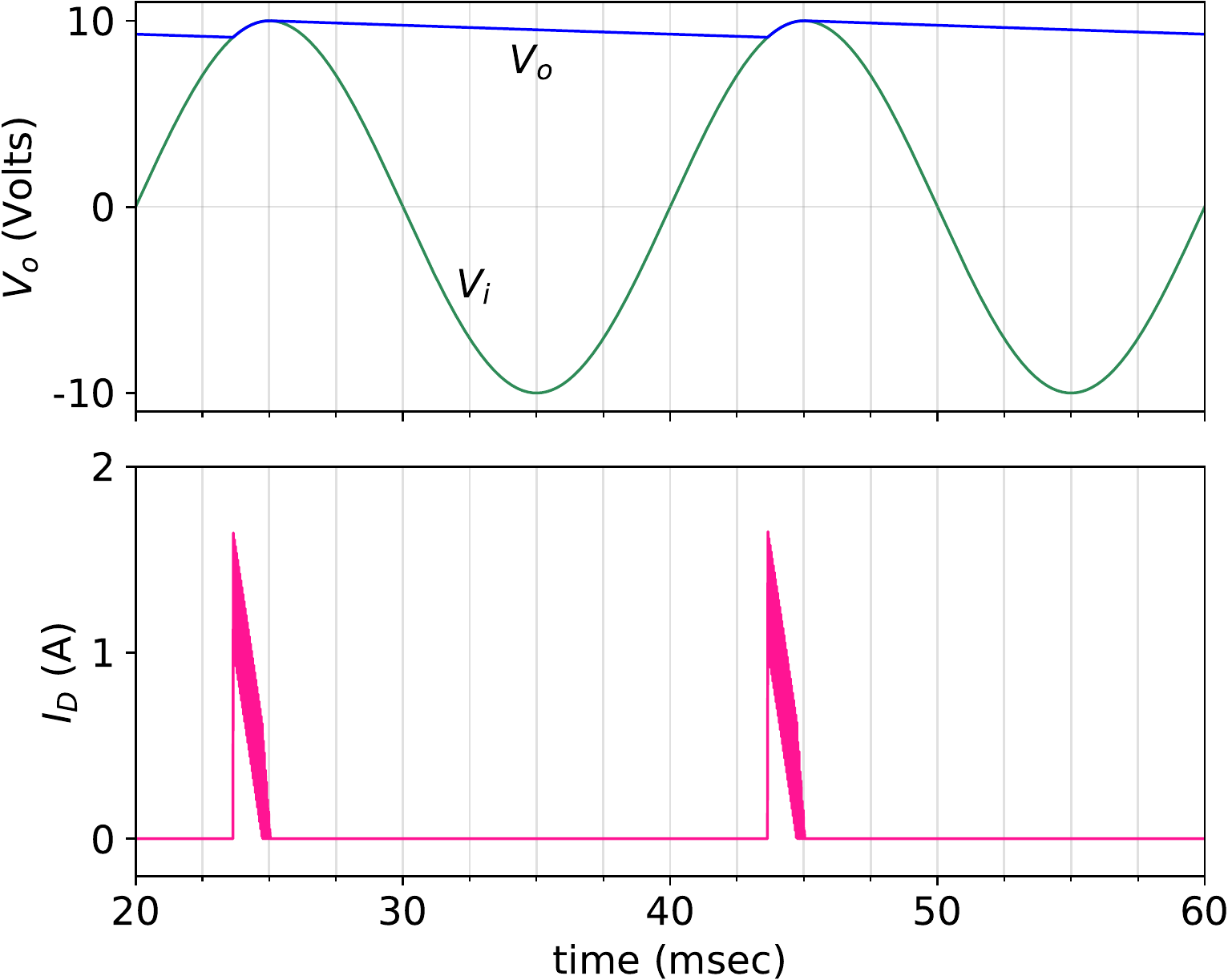}}
\vspace*{-0.7cm}
\caption{Input and output voltages, and diode current obtained with the FE
method for the half-wave rectifier circuit of Fig.~\ref{fig_hwr_3} with time
step $h \,$=$\, 20\,\mu$sec.}
\label{fig_hwr_5}
\end{figure}

\begin{figure}[!ht]
\centering
{\includegraphics[width=0.49\textwidth]{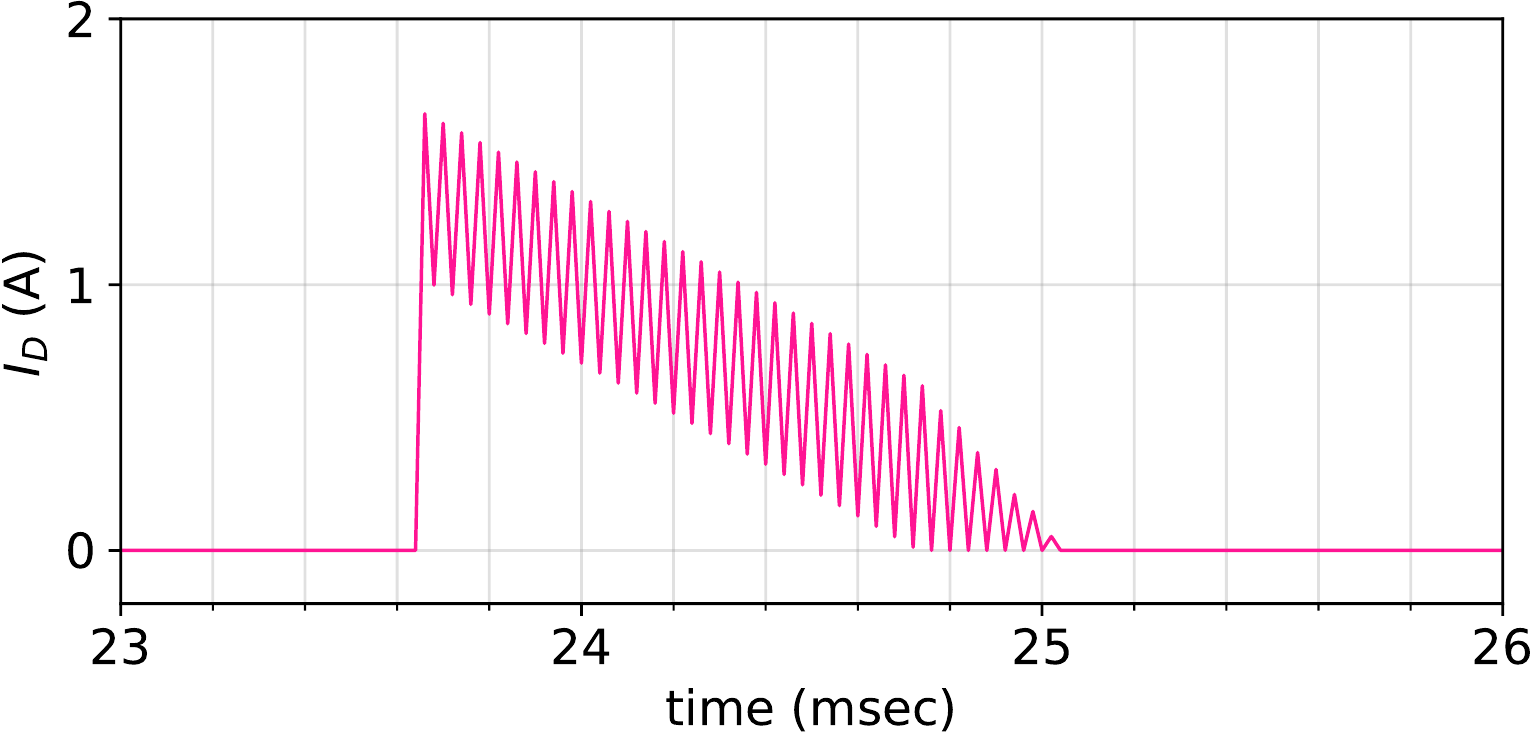}}
\vspace*{-0.7cm}
\caption{Diode current obtained with the FE
method for the half-wave rectifier circuit of Fig.~\ref{fig_hwr_3} with time
step $h \,$=$\, 20\,\mu$sec.}
\label{fig_hwr_6}
\end{figure}

The stability problem is illustrated in
Figs.~\ref{fig_hwr_4} and
\ref{fig_hwr_5}.
When $h < 20\,\mu$sec is used (Fig.~\ref{fig_hwr_4}), the
simulation results are as expected.
For $h \,$=$\, 20\,\mu$sec (Figs.~\ref{fig_hwr_5} and \ref{fig_hwr_6}), oscillations can be
seen in the diode current, indicating unstable behaviour of the FE method.

\subsection{Runge-Kutta-Fehlberg (RKF) method}
\label{sec_hwrect_rkf}
The RKF method is an adaptive (variable) time step method in which a
Runge-Kutta method of order four is used together with a Runge-Kutta method
of order five to obtain an estimate of the local truncation error (LTE) in
a given time step. If the LTE is smaller than a user-specified tolerance,
the next time step is allowed to be larger than the current time step.
On the other hand, if the LTE is larger than the tolerance, then the current
time step is rejected, and a smaller time step is tried. The RKF method requires
more work at each time step (six function evaluations) as compared to the FE
method (one function evaluation). However, its auto time step feature allows a
dramatic reduction in the total number of time points required for the simulation
while simultaneously satisfying the LTE constraint, and it is therefore much more
efficient than the FE method.

The RKF method for a single ordinary differential equation (ODE)
$\displaystyle\frac{dx}{dt} \,$=$\, f(t,x)$
can be described as follows. First, compute
$k_1$, $k_2$, $\cdots$, $k_6$ as,
\begin{eqnarray}
 k_1
 \!\!\! &=& \!\!\!
 h\,f(t_n, x_n),
 \label{eq_rkf_k1}
 \\
 k_2
 \!\!\! &=& \!\!\!
 h\,f\left(t_n + \alpha _2 h, x_n + \beta _{2,1} \,k_1\right),
 \label{eq_rkf_k2}
 \\
 k_3
 \!\!\! &=& \!\!\!
 h\,f(t_n + \alpha _3 h, x_n + \sum _{j=1}^{2} \beta _{3,j}\,k_j),
 \\
 k_4
 \!\!\! &=& \!\!\!
 h\,f(t_n + \alpha _4 h, x_n + \sum _{j=1}^{3} \beta _{4,j}\,k_j),
 \\
 k_5
 \!\!\! &=& \!\!\!
 h\,f(t_n + \alpha _5 h, x_n + \sum _{j=1}^{4} \beta _{5,j}\,k_j),
 \\
 k_6
 \!\!\! &=& \!\!\!
 h\,f(t_n + \alpha _6 h, x_n + \sum _{j=1}^{5} \beta _{6,j}\,k_j),
\end{eqnarray}
where $\alpha _i \leq 1$. The fourth- and fifth-order estimates for
$x_{n+1}$ are then given by
\begin{eqnarray}
 x_{n+1}\,(4^{\mathrm{th}}~{\textrm{order}})
 \!\!\! &=& \!\!\!
 x_n + \sum _{j=1}^{5} \gamma _j^{(4)}\,k_j,
 \label{eq_rkf_4th}
 \\
 x_{n+1}\,(5^{\mathrm{th}}~{\textrm{order}})
 \!\!\! &=& \!\!\!
 x_n + \sum _{j=1}^{6} \gamma _j^{(5)}\,k_j.
 \label{eq_rkf_5th}
\end{eqnarray}
The constants $\alpha$, $\beta$, $\gamma$ in the above equations can
be found in \cite{burden}. The difference between the fourth- and fifth-order
$x_{n+1}$ values
gives an estimate of the LTE. If the LTE is smaller than a tolerance, then
the fifth order $x_{n+1}$ value
is accepted as the numerical solution of the ODE at $t_{n+1}$; if not, the
current time step is rejected, and a revised value of $h$ is computed, using
the LTE estimate\,\cite{burden}.

For the half-wave rectifier of Fig.~\ref{fig_hwr_3}, the ODE comes from the
capacitor and is given by
\begin{equation}
\displaystyle\frac{dV_c}{dt} = \displaystyle\frac{1}{C}\,i_3
\label{eq_hwrect_rkf_1}
\end{equation}
(see Fig.~\ref{fig_hwr_3}).
The first step in applying the RKF method to this problem is to compute
$k_1$ (see Eq.~\ref{eq_rkf_k1}) as
\begin{equation}
k_1 = h\,\displaystyle\frac{1}{C}\,i_{3,n},
\label{eq_hwrect_rkf_2}
\end{equation}
where $i_{3,n}$ is the numerical solution for $i_3$ at $t_n$ and is already
known. For the next RKF step (Eq.~\ref{eq_rkf_k2}), we compute an intermediate
value of $i_3$ (denoted by $i_3^{(2)}$) using the following steps.
\begin{list}{(\roman{cntr2})}{\usecounter{cntr2}}
 \item
  Update $V_C^{(2)}$ as
\begin{equation}
  V_C^{(2)} = V_{C,n} + \beta _{2,1}\,k_1.
\label{eq_hwrect_rkf_3}
\end{equation}
 \item
  Update $V_s^{(2)}$ as
\begin{equation}
  V_s^{(2)} = V_m\,\sin \omega\,(t_n + \alpha _2\,h).
\label{eq_hwrect_rkf_4}
\end{equation}
 \item
  Solve the following set of equations, making sure
  that the {\bf{A}} matrix (corresponding to diode on or off)
  used for solving the equations is consistent with the solution.
\begin{equation}
\label{eq_hwrect_rkf_5_1}
i_1^{(2)} + i_2^{(2)} = 0,
\end{equation}
\begin{equation}
-i_2^{(2)} + i_3^{(2)} + i_5^{(2)} = 0,
\label{eq_hwrect_rkf_5_2}
\end{equation}
\begin{equation}
-i_3^{(2)} + i_4^{(2)} = 0,
\label{eq_hwrect_rkf_5_3}
\end{equation}
\begin{equation}
V_1^{(2)} = V_s^{(2)}
\label{eq_hwrect_rkf_5_4}
\end{equation}
\begin{equation}
V_2^{(2)} - V_3^{(2)} = V_C^{(2)},
\label{eq_hwrect_rkf_5_5}
\end{equation}
\begin{equation}
i_4^{(2)} - G_1 V_3^{(2)} = 0,
\label{eq_hwrect_rkf_5_6}
\end{equation}
\begin{equation}
i_5^{(2)} - G V_2^{(2)} = 0,
\label{eq_hwrect_rkf_5_7}
\end{equation}
\begin{equation}
\begin{array}{cl}
V_1^{(2)} - V_2^{(2)} = 0 &{\textrm{if}}~D~{\textrm{is on}}, \\
i_2^{(2)} = 0 &{\textrm{if}}~D~{\textrm{is off}}.
\end{array}
\label{eq_hwrect_rkf_5_8}
\end{equation}
\end{list}
We are now in a position to compute $k_2$ and $V_C^{(3)}$ as
\begin{equation}
k_2 = h\,\displaystyle\frac{1}{C}\,i_3^{(2)},
\label{eq_hwrect_rkf_6}
\end{equation}
\begin{equation}
V_C^{(3)} =
V_C^{(2)}
+ \beta _{3,1}\,k_1
+ \beta _{3,2}\,k_2.
\label{eq_hwrect_rkf_7}
\end{equation}
We then compute
$V_s^{(3)}$
as
$V_m\,\sin \omega\,(t_n + \alpha _3\,h)$, solve the circuit
equations to obtain
$i_3^{(3)}$, and compute $k_3$.
Proceeding in this manner, we compute
$k_1$, $k_2$, $\cdots$, $k_6$, and finally
the fourth- and fifth-order estimates for $V_{C,n+1}$
(Eqs.~\ref{eq_rkf_4th} and \ref{eq_rkf_5th}).
If there are several capacitors and inductors in the circuit,
we would evaluate
Eq.~\ref{eq_hwrect_rkf_3} for each of the state variables
(inductor currents and capacitor voltages), update source value(s),
and then solve the electrical equations.

\begin{figure}[!ht]
\centering
{\includegraphics[width=0.49\textwidth]{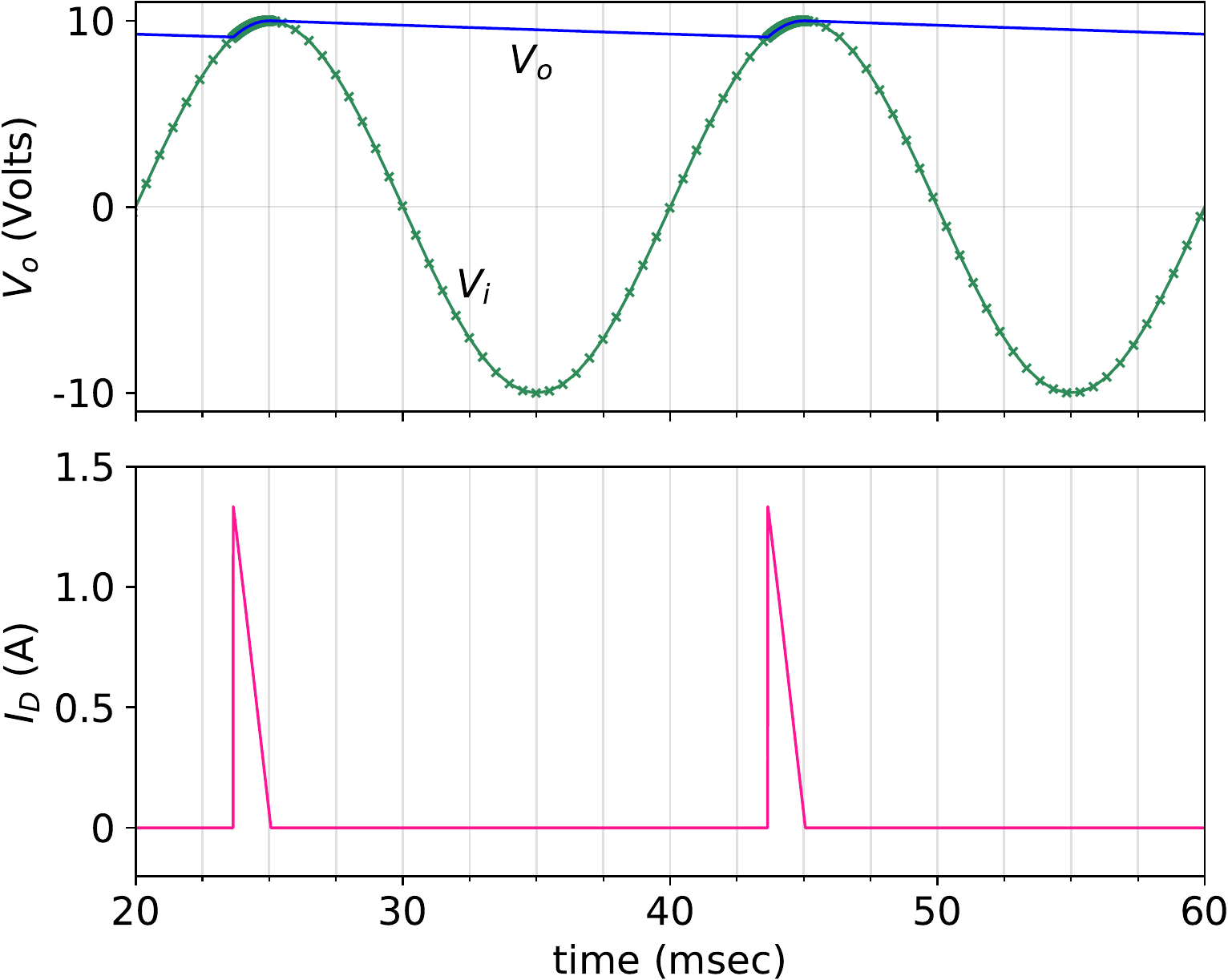}}
\vspace*{-0.7cm}
\caption{Input and output voltages, and diode current obtained with the RKF
method for the half-wave rectifier circuit of Fig.~\ref{fig_hwr_3}.
The simulation time points are shown by crosses in the $V_i$ plot.}
\label{fig_hwr_7}
\end{figure}

Fig.~\ref{fig_hwr_7} shows the results obtained with the RKF method.
The time points used by the RKF method are shown by crosses in the
$V_i(t)$ plot. It can be seen that, in the charging phase (where the time
constant is small), small time steps are taken by the RKF method.
In the discharging phase, the time constant is large, and much larger time
steps~-- limited in this case by a user-specified maximum time step of
0.5\,msec~-- are taken by the RKF method, thus leading to a much smaller
number of total time points, as compared to the FE method.

The half-wave rectifier example brings out the fundamental
limitation of using explicit methods for electrical circuits,
viz., the necessity of using small time steps because of stability
constraints. However, in several power electronic
circuits of interest, the ELEX scheme essentially bypasses the
stability issue by avoiding small time constants altogether, and
in that case, the time step is limited by accuracy, rather than
stability.

\section{Switches in series}
\label{sec_sw}
\begin{figure}[!ht]
\centering
\scalebox{0.9}{\includegraphics{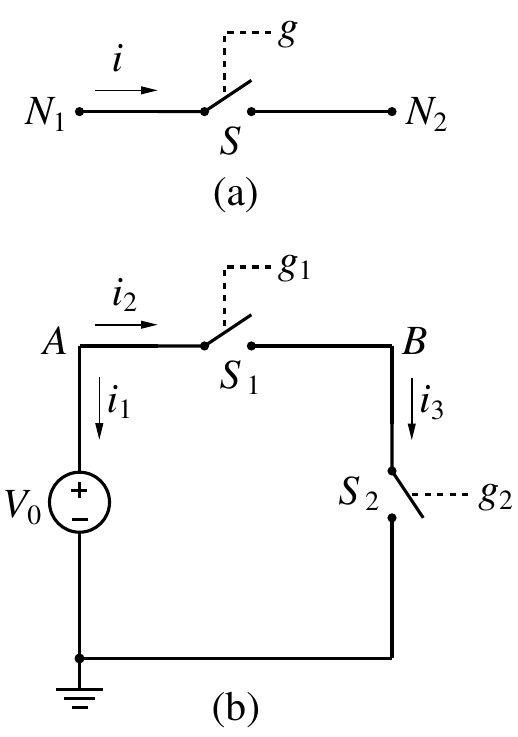}}
\vspace*{-0.2cm}
\caption{(a)\,Schematic diagram of a controlled switch,
(b)\,A circuit with two controlled switches in series.}
\label{fig_sw_1}
\end{figure}

\begin{figure}[!ht]
\centering
{\includegraphics[width=0.49\textwidth]{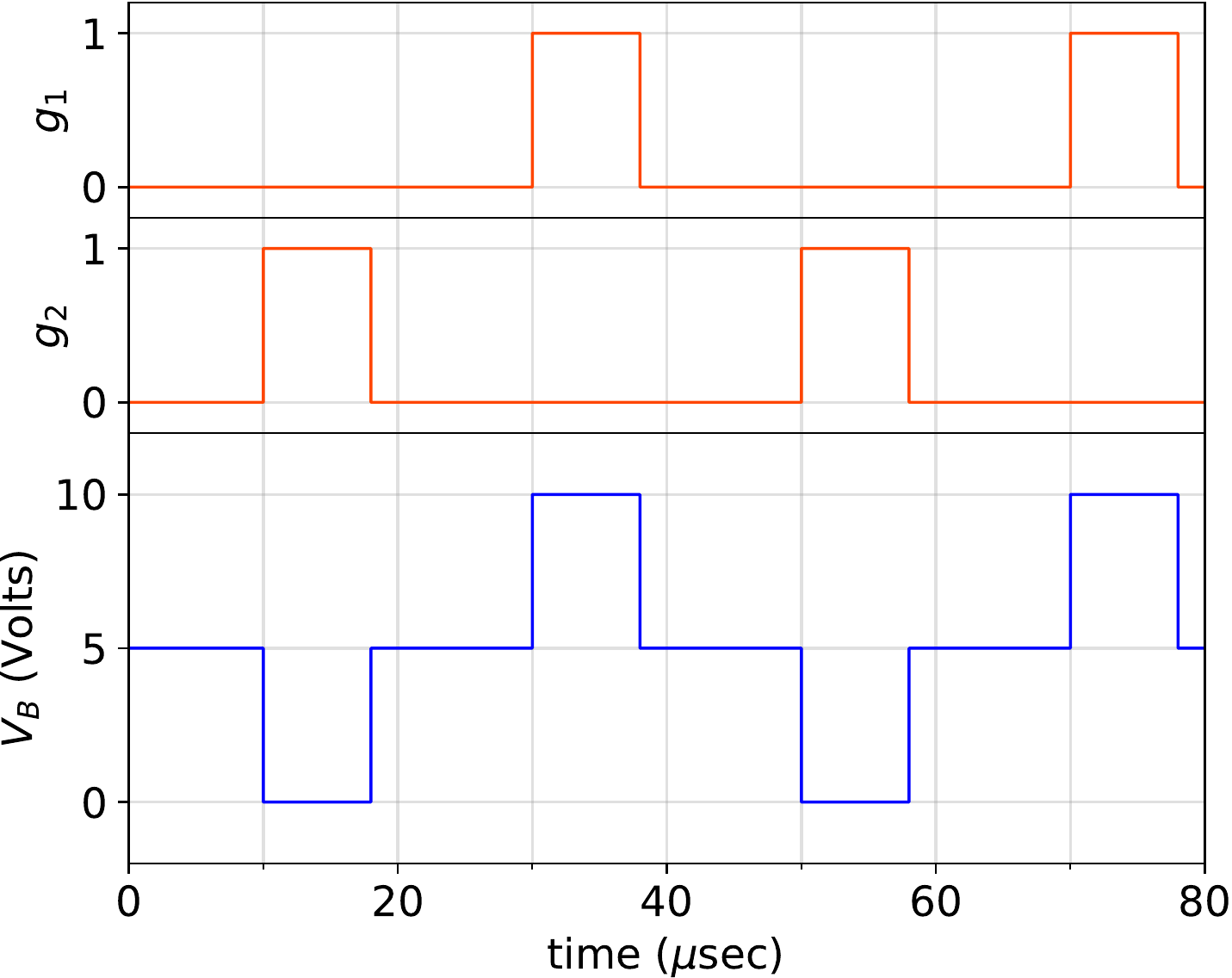}}
\vspace*{-0.7cm}
\caption{Sample waveforms expected for
the circuit of Fig.~\ref{fig_sw_1}\,(b) with $V_0 \,$=$\, 10$\,V.}
\label{fig_sw_2}
\end{figure}

Consider the controlled switch, e.g., an idealised MOS transistor,
shown in
Fig.~\ref{fig_sw_1}\,(a).
In the ELEX scheme, we can write the switch equation as
\begin{equation}
\begin{array}{cl}
V_{N1} - V_{N2} = 0 &{\textrm{if}}~D~{\textrm{is on}}, \\
i = 0 &{\textrm{if}}~D~{\textrm{is off}}.
\end{array}
\label{eq_sw_1}
\end{equation}
In most situations, the above approach would work fine. However,
this simple model can fail for some specific situations. As an example,
consider the circuit shown in
Fig.~\ref{fig_sw_1}\,(b)
in which $V_0$ is a dc (known) voltage.
If $g_1 \,$=$\, 1$ and
$g_2 \,$=$\, 0$, we expect $V_B \,$=$\, V_0$.
If $g_1 \,$=$\, 0$ and
$g_2 \,$=$\, 1$, we expect $V_B \,$=$\, 0$
(see Fig.~\ref{fig_sw_2}).
When both $g_1$ and $g_2$ are zero, we would expect the applied
voltage $V_0$ to split equally between the two switches, and $V_B$
should therefore be 5\,V, as shown in Fig.~\ref{fig_sw_2}.
Let us now apply the ELEX scheme to this circuit.

Once again, we write the equations in a systematic manner, writing
the KCL equations first, followed by the element equations, and get
\begin{equation}
i_1 + i_2 = 0,
\label{eq_sw_2_1}
\end{equation}
\begin{equation}
i_2 - i_3 = 0,
\label{eq_sw_2_2}
\end{equation}
\begin{equation}
V_A = V_0,
\label{eq_sw_2_3}
\end{equation}
\begin{equation}
\begin{array}{cl}
V_A - V_B = 0 &{\textrm{if}}~g_1 = 1, \\
i_2 = 0 &{\textrm{if}}~g_1 = 0,
\end{array}
\label{eq_sw_2_4}
\end{equation}
\begin{equation}
\begin{array}{cl}
V_B = 0 &{\textrm{if}}~g_2 = 1, \\
i_3 = 0 &{\textrm{if}}~g_2 = 0,
\end{array}
\label{eq_sw_2_5}
\end{equation}
where we have applied Eq.~\ref{eq_sw_1} to each of the two switches.

If one of $g_1$ and $g_2$ is high,
Eqs.~\ref{eq_sw_2_1}-\ref{eq_sw_2_5}
give us the expected result.
For the case $g_1 \,$=$\, g_2 \,$=$\, 0$, no solution is possible
because node $B$ becomes isolated. In other words, if the equations
are written in the form
${\bf{A}}\,{\bf{x}} \,$=$\, {\bf{b}}$,
then {\bf{A}} turns out to be singular. Although this situation is
rare in power electronic circuits, a general-purpose simulation program
must be able to handle it suitably.

When both $S_1$ and $S_2$ are off, we expect $V_0$ to divide equally
between them, which suggests that the switch should be modelled as
a large resistance
$R_{\mathrm{off}}$
in the off state. However, in the ELEX scheme, we would like to
avoid such a representation because of its potential to create
small $L/R$-type time constants. What we need is a procedure that
allows voltage division between the switches, but without replacing
the switch with
$R_{\mathrm{off}}$.
One way to achieve these conflicting goals is as follows.

\begin{figure}[!ht]
\centering
\scalebox{0.9}{\includegraphics{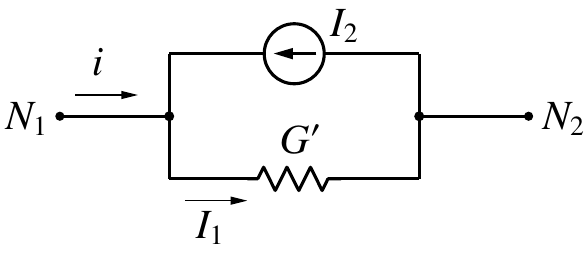}}
\vspace*{-0.2cm}
\caption{Equivalent circuit of a controlled switch in the off state
for implementation in the ELEX scheme.}
\label{fig_sw_3}
\end{figure}

When the gate voltage is high, we continue to model the switch
with the ideal switch equation
\begin{equation}
V_{N1} - V_{N2} = 0.
\label{eq_sw_3}
\end{equation}
When the gate voltage is low, we use the model shown in
Fig.~\ref{fig_sw_3}, where
$G' \,$=$\, 1/R_{\mathrm{off}}$
is a small conductance.

The current $I_2$ is given by,
\begin{equation}
I_2 = \displaystyle\frac{k-1}{k_{\mathrm{max}}-1}\,G'\,(V_{N1}-V_{N2}),
\label{eq_sw_4}
\end{equation}
where $k$ takes values, 1, 2, $\cdots$, $k_{\mathrm{max}}$. The switch current
$i$ in the off state is then given by
\begin{equation}
i - G'\,(V_{N1}-V_{N2}) = -\, \displaystyle\frac{k-1}{k_{\mathrm{max}}-1}\,G'\,(V_{N1}-V_{N2}).
\label{eq_sw_5}
\end{equation}

With $k \,$=$\, 1$, Eq.~\ref{eq_sw_5} amounts to removing the current source
$I_2$ in Fig.~\ref{fig_sw_3}, keeping only $G'$. When the circuit equations
are solved for this condition, voltages get suitably assigned. For the example
of Fig.~\ref{fig_sw_1}\,(b), we would get $V_B \,$=$\, V_0/2$.
When 
$k \,$=$\, k_{\mathrm{max}}$,
Eq.~\ref{eq_sw_5} reduces to $i \,$=$\, 0$, the ideal switch equation in the
off state, since $I_1$ and $I_2$ become equal. Thus, by increasing $k$ from 1 to
$k_{\mathrm{max}}$,
we achieve our objective of establishing appropriate node voltages without
replacing the switch with
$R_{\mathrm{off}}$.
For the circuit of of Fig.~\ref{fig_sw_1}\,(b),
$k_{\mathrm{max}} \,$=$\, 2$
is adequate. It remains to be seen if a larger value of
$k_{\mathrm{max}}$
is necessary for some other circuits.

Eq.~\ref{eq_sw_5}, for a given value of $k$, is solved in an iterative
manner, treating
$V_{N1}$,
$V_{N2}$
on the right-hand side (RHS) as known values coming from the previously
computed solution. The RHS becomes part of {\bf{b}} in
${\bf{A}}\,{\bf{x}} \,$=$\, {\bf{b}}$,
the complete set of circuit equations.
Solving
${\bf{A}}\,{\bf{x}} \,$=$\, {\bf{b}}$,
we obtain revised values of
$V_{N1}$,
$V_{N2}$,
and use those to compute the RHS of Eq.~\ref{eq_sw_5} again, and so on
until convergence. For the circuit of
Fig.~\ref{fig_sw_1}\,(b),
two iterations suffice.

Clearly, the above procedure comes with an additional computational
cost and should therefore be used only if the ideal switch model
(Eq.~\ref{eq_sw_1}) is creating a singular matrix situation. In most
power electronic circuits, the ideal switch model seems to work well.

\section{Boost converter}
\label{sec_boost}
\begin{figure}[!ht]
\centering
\scalebox{0.9}{\includegraphics{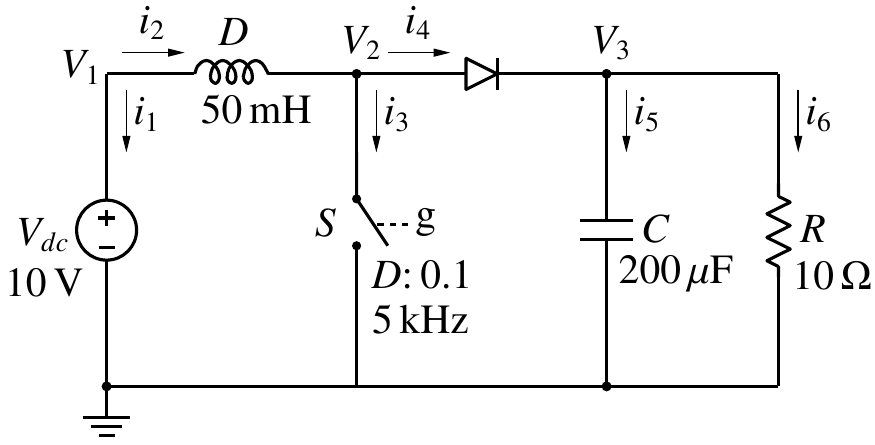}}
\vspace*{-0.2cm}
\caption{Boost converter circuit.}
\label{fig_boost_1}
\end{figure}

\begin{figure}[!ht]
\centering
{\includegraphics[width=0.49\textwidth]{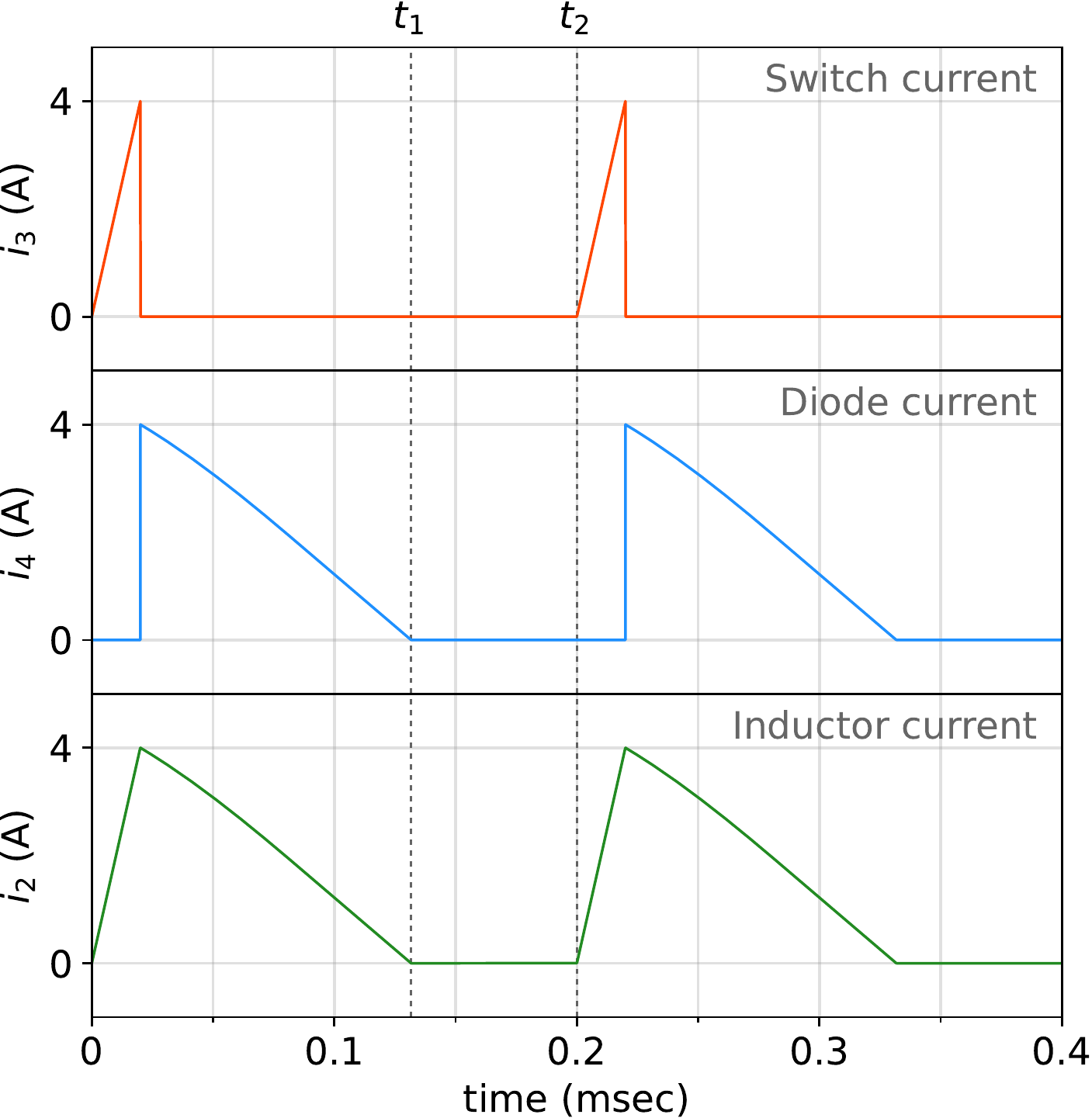}}
\vspace*{-0.7cm}
\caption{Steady-state waveforms for the boost converter circuit
of Fig.~\ref{fig_boost_1}.}
\label{fig_boost_2}
\end{figure}

Consider the boost converter circuit shown in
Fig.~\ref{fig_boost_1}.
For the parameter values given in the figure, the inductor
current (in the steady state) is discontinuous, as shown in
Fig.~\ref{fig_boost_2}. The purpose of this section is to
describe the challenges presented by the discontinuous nature
of the inductor current and to present a way to address the same.

The ELEX-RKF procedure to obtain the solution at $t_{n+1}$~--
assuming as always the solution at $t_n$ to be already
available~-- starts with
\begin{equation}
k_1^{(C)} = h\,\displaystyle\frac{1}{C}\,i_{5,n},
\label{eq_boost_1}
\end{equation}
\begin{equation}
k_1^{(L)} = h\,\displaystyle\frac{1}{L}\,V_{L,n},
\label{eq_boost_2}
\end{equation}
for the capacitor and inductor, respectively. The inductor voltage
$V_L$ for this circuit is $V_L \,$=$\, V_1-V_2$.

Next, we obtain
$V_C^{(2)}$ and
$i_L^{(2)}$,
the phase-2 updates for $V_C$ and $i_L$, as
\begin{equation}
V_C^{(2)} = V_{C,n} + \beta _{2,1} \,k_1^{(C)},
\label{eq_boost_3}
\end{equation}
\begin{equation}
i_L^{(2)} = i_{L,n} + \beta _{2,1} \,k_1^{(L)},
\label{eq_boost_4}
\end{equation}
update the gate voltage value for the MOS switch, and then solve
the set of circuit equations given by
\begin{equation}
i_1^{(2)} + i_2^{(2)} = 0,
\label{eq_boost_elex_first}
\end{equation}
\begin{equation}
i_2^{(2)} - i_3^{(2)} - i_4^{(2)} = 0,
\end{equation}
\begin{equation}
i_4^{(2)} - i_5^{(2)} - i_6^{(2)} = 0,
\end{equation}
\begin{equation}
V_1^{(2)} = V_{dc},
\end{equation}
\begin{equation}
i_2^{(2)} = i_L^{(2)},
\label{eq_il}
\end{equation}
\begin{equation}
V_3^{(2)} = V_C^{(2)},
\end{equation}
\begin{equation}
i_6^{(2)} - G\,V_3^{(2)} = 0,
\end{equation}
\begin{equation}
i_4^{(2)} - i_d^{(2)} = 0,
\end{equation}
\begin{equation}
i_3^{(2)} - i_{sw}^{(2)} = 0,
\end{equation}
\begin{equation}
\begin{array}{cl}
V_2^{(2)} - V_3^{(2)} = V_{\mathrm{on}} &{\textrm{if}}~D~{\textrm{is on}}, \\
\T i_d^{(2)} = 0 &{\textrm{if}}~D~{\textrm{is off}},
\end{array}
\end{equation}
\begin{equation}
\begin{array}{cl}
V_2^{(2)} = 0 &{\textrm{if}}~S~{\textrm{is on}}, \\
\T i_{sw}^{(2)} = 0 &{\textrm{if}}~S~{\textrm{is off}}.
\end{array}
\label{eq_boost_elex_last}
\end{equation}
Having obtained the solution at stage 2 of the RKF process, we then proceed
to compute
$k_2^{(C)}$,
$k_2^{(L)}$,
solve the circuit equations to obtain the solution at stage 3, and so on.
In a way, this is a straightforward extension of the ELEX-RKF scheme we have
seen in
Sec.~\ref{sec_hwrect_rkf}. However, one major change is required.

In the interval from $t_1$ to $t_2$ in Fig.~\ref{fig_boost_2}, both $S$ and $D$
are off, i.e.,
$i_3 \,$=$\, i_4 \,$=$\, 0$. With this condition, the
${\bf{A}}\,{\bf{x}} \,$=$\, {\bf{b}}$
problem described by
Eqs.~\ref{eq_boost_elex_first}-\ref{eq_boost_elex_last}
has no solution because {\bf{A}} becomes singular. Intuitively, we would
expect that to happen since we are trying to assign a value
to the inductor current when there is no path for that current.
Even a boost converter which has continuous conduction in the steady
state can go through the above situation (where both $D$ and $S$ are
simultaneously off) in the transient phase, i.e., before steady state
is attained by the circuit.

\begin{figure}[!ht]
\centering
\scalebox{0.9}{\includegraphics{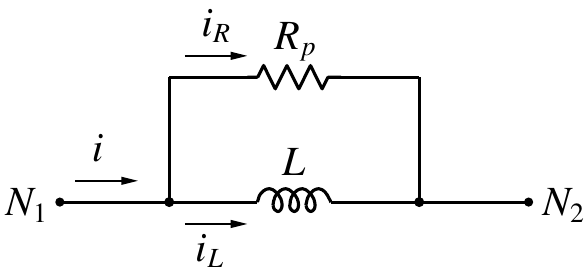}}
\vspace*{-0.2cm}
\caption{Addition of a resistance in parallel with an inductor to address
the singular matrix issue (see text).}
\label{fig_boost_3}
\end{figure}

The singular matrix issue can be addressed by connecting a resistance $R_p$
in parallel with $L$ as shown in Fig.~\ref{fig_boost_3}.
The $R_p$ value should be chosen to be sufficiently large (say, 1\,M$\Omega$)
such that it draws a negligibly small current, thus leaving the circuit
solution essentially undisturbed. With $R_p$ in place,
the inductor current equation (Eq.~\ref{eq_il}) gets replaced by
\begin{equation}
i_2^{(2)} - \left(V_1^{(2)} - V_2^{(2)}\right)/R_p = i_L^{(2)},
\end{equation}
and the circuit matrix {\bf{A}}
remains non-singular even if $i_3$ and $i_4$ in Fig.~\ref{fig_boost_1} become zero.

The turn-off process of the diode at time $t_1$ (see Fig.~\ref{fig_boost_2})
needs to be carefully handled for the following reason. The inductor current
in the ELEX-RKF scheme gets updated in several stages of the RKF process, the
first step being
\begin{equation}
i_L^{(2)} = i_{L,n} + \beta _{2,1} \,\displaystyle\frac{h}{L}\,V_{L,n},
\label{eq_boost_il_1}
\end{equation}
where
$i_{L,n}$ and
$V_{L,n}$
come from the solution obtained at $t \,$=$\, t_n$. It is crucial to catch the
diode turn-off point as closely as possible, i.e., the diode current should be
sufficiently small before we move on to the off condition of the diode.
This is achieved in the ELEX scheme using binary search~-- as in the
PLECS\,\cite{plecs} program~-- where we keep narrowing the bracket around the
exact turn-off point, until it becomes acceptably small, say,
$\Delta t \,$=$\, 10^{-12}$\,sec. When that is achieved, we have two simulation
time points, one before and one after the turn-off event, the interval between
them being $\Delta t$. Let us denote these two points by
$t_{\mathrm{on}}'$ and
$t_{\mathrm{off}}'$.
At $t_{\mathrm{on}}'$, the diode is still on, but $i_d$ ($i_4$ in Fig.~\ref{fig_boost_1})
is small, say, $10^{-12}$\,A or smaller.
At $t_{\mathrm{off}}'$, the diode has turned off, $i_d$ is exactly zero, and
KCL implies that
$i_2$ (in Fig.~\ref{fig_boost_1})
is also zero, since the MOS switch is already off.
The currents $i_L$ and $i_R$ in
Fig.~\ref{fig_boost_3}
at $t \,$=$\, t_{\mathrm{off}}'$ cancel each other exactly. These currents
are individually small but not zero. Considering this situation, in order to
ensure that $i_L$ goes to zero quickly, we can use
\begin{equation}
i_L^{(2)} = i_{2,n} + \beta _{2,1} \,\displaystyle\frac{h}{L}\,V_{L,n},
\label{eq_boost_il_2}
\end{equation}
rather than Eq.~\ref{eq_boost_il_1}. Note that
$i_{L,n}$ has been replaced with
$i_{2,n}$ on the right-hand side in
Eq.~\ref{eq_boost_il_2}.

With the above modification, the ELEX-RKF scheme was able to handle the
boost converter circuit of Fig.~\ref{fig_boost_1} accurately. It was
observed that, if the diode current at
$t_{\mathrm{on}}'$
is not small enough, the inductor current does not reduce to zero. As we
proceed to the off state of the diode, we have a situation in which both
the diode and MOS switch are off, and $i_L \,$=$\, -\,i_R$ (Fig.~\ref{fig_boost_3}).
In other words, current circulates in the $L$-$R_p$ loop.
This is a disastrous situation because the small time constant
$L/R_p$ of the $L$-$R_p$ loop forces the RKF process to take very
small time steps, and the simulation becomes impractically slow.

In a general scenario, the $L$-$R_p$ combination is useful in another
sense. If a substantial inductor current is forced to become zero
abruptly, the current would then go through $R_p$, causing an
unrealistically large voltage drop across the inductor. This voltage
drop can be monitored by the simulator to issue a suitable warning or
error message.

\section{Closed-loop circuits}
\label{sec_cloop}
\begin{figure}[!ht]
\centering
\scalebox{0.9}{\includegraphics{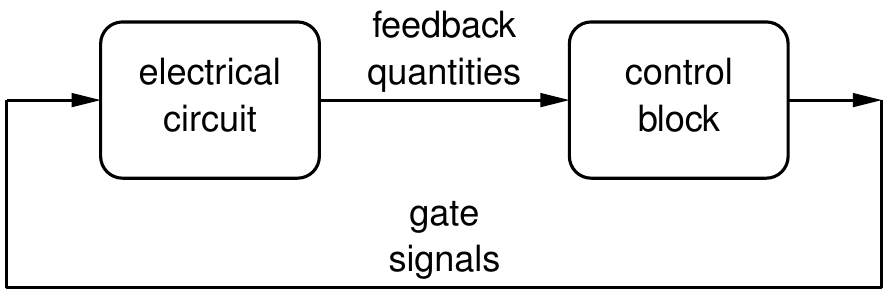}}
\vspace*{-0.2cm}
\caption{Typical structure of a closed-loop power electronic circuit.}
\label{fig_cloop_1}
\end{figure}

In a closed-loop power electronic circuit, additional variables related
to the control block are involved.
Typically, the electrical circuit produces one or more
feedback quantities (currents or voltages), which serve as inputs
to the control block, as shown in Fig.~\ref{fig_cloop_1}.
The control block in turn generates the gate signals required by the
electrical circuit. In general, the control block is easier to handle
than the electrical block because it involves only a ``flow graph"
and not a network where KCL and KVL equations need to be satisfied.
When an explicit method is used to handle the control block, as we
would be doing in the ELEX-RKF scheme, several ``passes" are made
to treat the elements of the control block, as dictated by the flow
graph (see
\cite{sequelmbp},
\cite{gseimmanual}).
In each pass, only those elements whose input values have been updated
are processed. In the following, we will consider two examples to illustrate
how the ELEX-RKF scheme can be applied to closed-loop circuits.

\subsection{Current-controlled two-level voltage source converter}
\label{sec_cloop_cc}
\begin{figure*}[!ht]
\centering
\scalebox{0.9}{\includegraphics{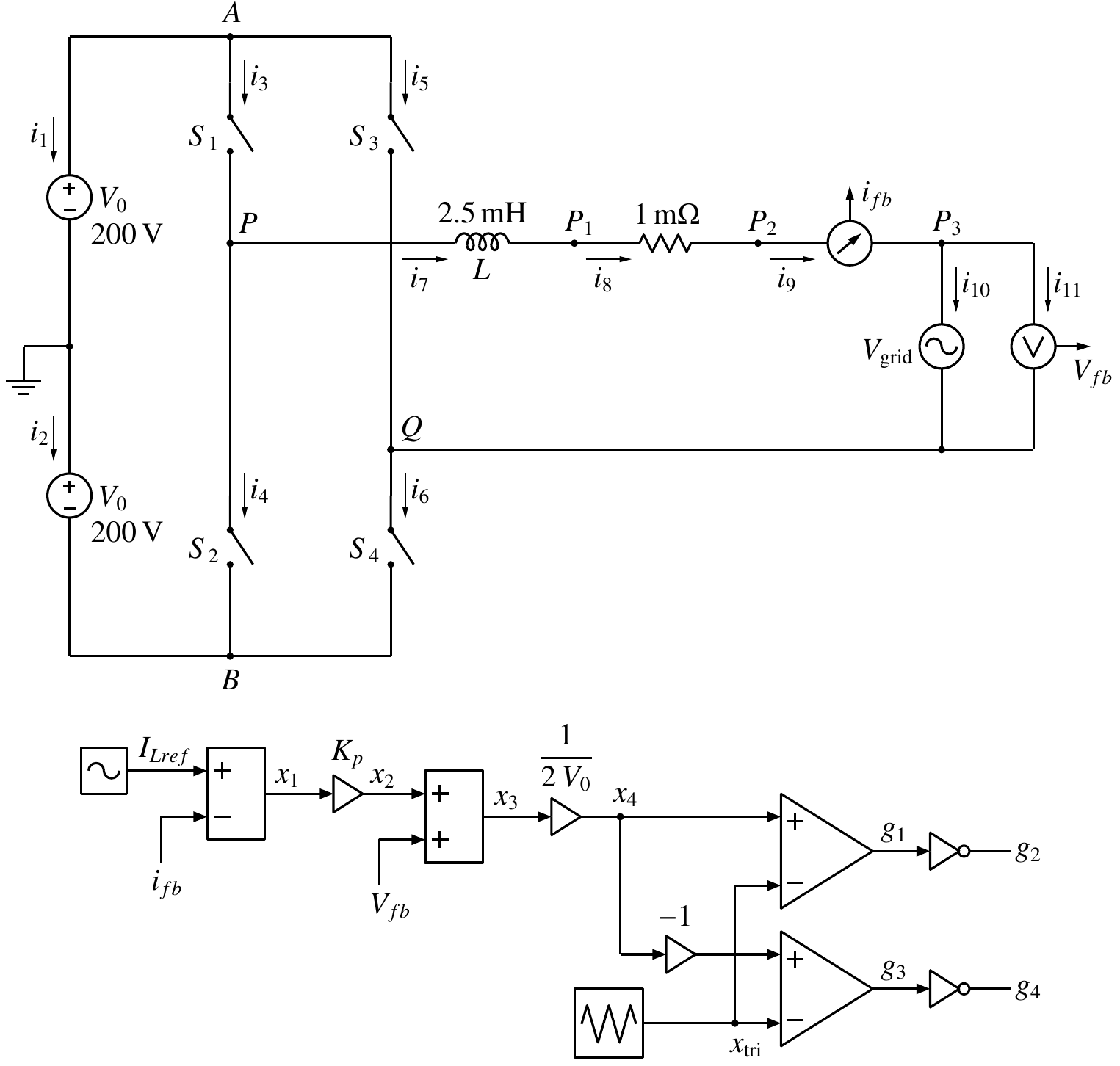}}
%\vspace*{-0.2cm}
\caption{Schematic diagram of current-controlled two-level voltage source
converter. The reference current is given by
$I_{Lref} \,$=$\, I_m\sin \,(\omega \,t+\phi)$,
where $I_m \,$=$\, 2.5\sqrt{2}\,$A,
$f \,$=$\, 50$\,Hz,
$\phi \,$=$\, -\pi/6$.
The triangular waveform is symmetric, with
minimum and maximum values of $-1$ and $1$, respectively, and
$f_{\mathrm{tri}} \,$=$\, 20$\,kHz.
$K_p$ is given by $K_p \,$=$\, \omega _{\mathrm{tri}}L/10$.}
\label{fig_cloop1_1}
\end{figure*}

The circuit diagram for a current-controlled two-level voltage source
converter is shown in Fig.~\ref{fig_cloop1_1}, along with the control
block.
We have already seen how voltage sources, MOS switches, resistors,
inductors, are treated in the ELEX-RKF scheme. Let us see how the two
new electrical elements in this circuit, viz., ammeter and voltmeter,
can be handled. The ammeter is like a dc voltage source with
$V_{dc} \,$=$\, 0$\,V, satisfying the equations,
\begin{equation}
V_{P2} - V_{P3} = 0.
\label{eq_cloop1_1}
\end{equation}
\begin{equation}
i_9 - i_{fb} = 0,
\label{eq_cloop1_2}
\end{equation}
The voltmeter is like a dc current source with
$I_{dc} \,$=$\, 0$\,A, satisfying the equations,
\begin{equation}
i_{11} = 0,
\label{eq_cloop1_3}
\end{equation}
\begin{equation}
V_{P3} - V_Q - V_{fb} = 0.
\label{eq_cloop1_4}
\end{equation}

The element equations, together with the KCL equations, constitute
the set of electrical equations which, as we have seen in
Sec.~\ref{sec_boost}, can be written in the form
${\bf{A}}\,{\bf{x}} \,$=$\, {\bf{b}}$.
Note that the feedback variables,
$i_{fb}$ and
$V_{fb}$,
are also included in {\bf{x}}.

At each stage of the RKF process, we solve the electrical equations
and obtain
$i_{fb}$,
$V_{fb}$.
This is followed by updating the ``control" variables
$I_{Lref}$,
$x_{\mathrm{tri}}$,
$x_1$,
$x_2$,
$x_3$,
$x_4$,
and finally the gate signals
$g_1$ to $g_4$.
In this example, the control block does not involve any state
variables.

The control variables are updated as follows.
First, we treat the source elements by assigning suitable values
to $I_{Lref}$ and
$x_{\mathrm{tri}}$.
The rest of the control variables are then updated as
\begin{equation}
x_1 = I_{Lref} - i_{fb},
\label{eq_cloop1_5_1}
\end{equation}
\begin{equation}
x_2 = K_p\,x_1,
\label{eq_cloop1_5_2}
\end{equation}
\begin{equation}
x_3 = x_2 - V_{fb},
\label{eq_cloop1_5_3}
\end{equation}
\begin{equation}
x_4 = \displaystyle\frac{1}{2\,V_0}\,x_3.
\label{eq_cloop1_5_4}
\end{equation}
The gate signals $g_1$ to $g_4$ are then obtained by comparing
$x_4$ or $-x_4$ with $x_{\mathrm{tri}}$.

\begin{figure}[!ht]
\centering
{\includegraphics[width=0.49\textwidth]{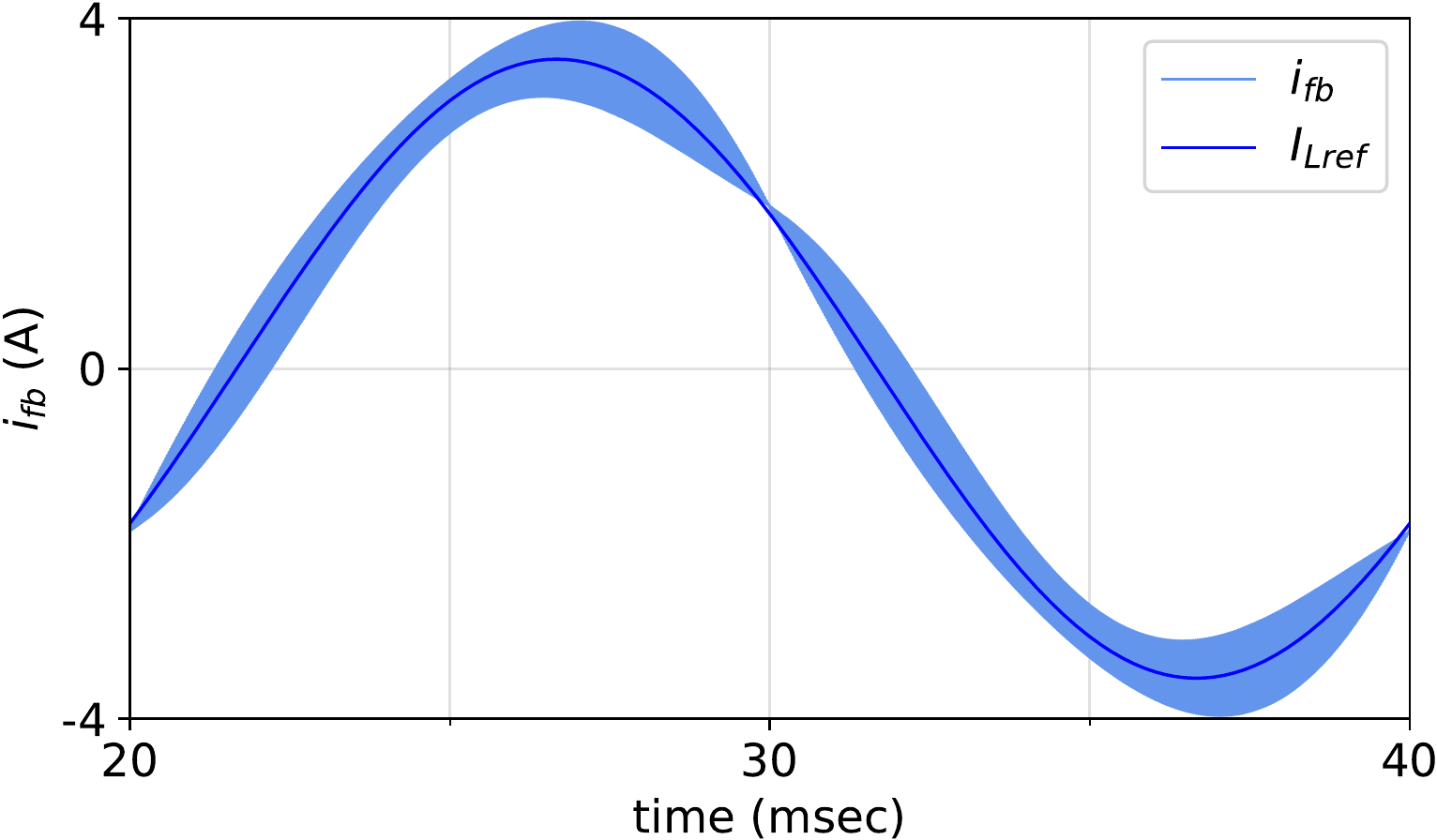}}
\vspace*{-0.7cm}
\caption{ELEX-RKF results for the current-controlled two-level voltage source
converter of Fig.~\ref{fig_cloop1_1}.}
\label{fig_cloop1_2}
\end{figure}

\begin{figure}[!ht]
\centering
{\includegraphics[width=0.49\textwidth]{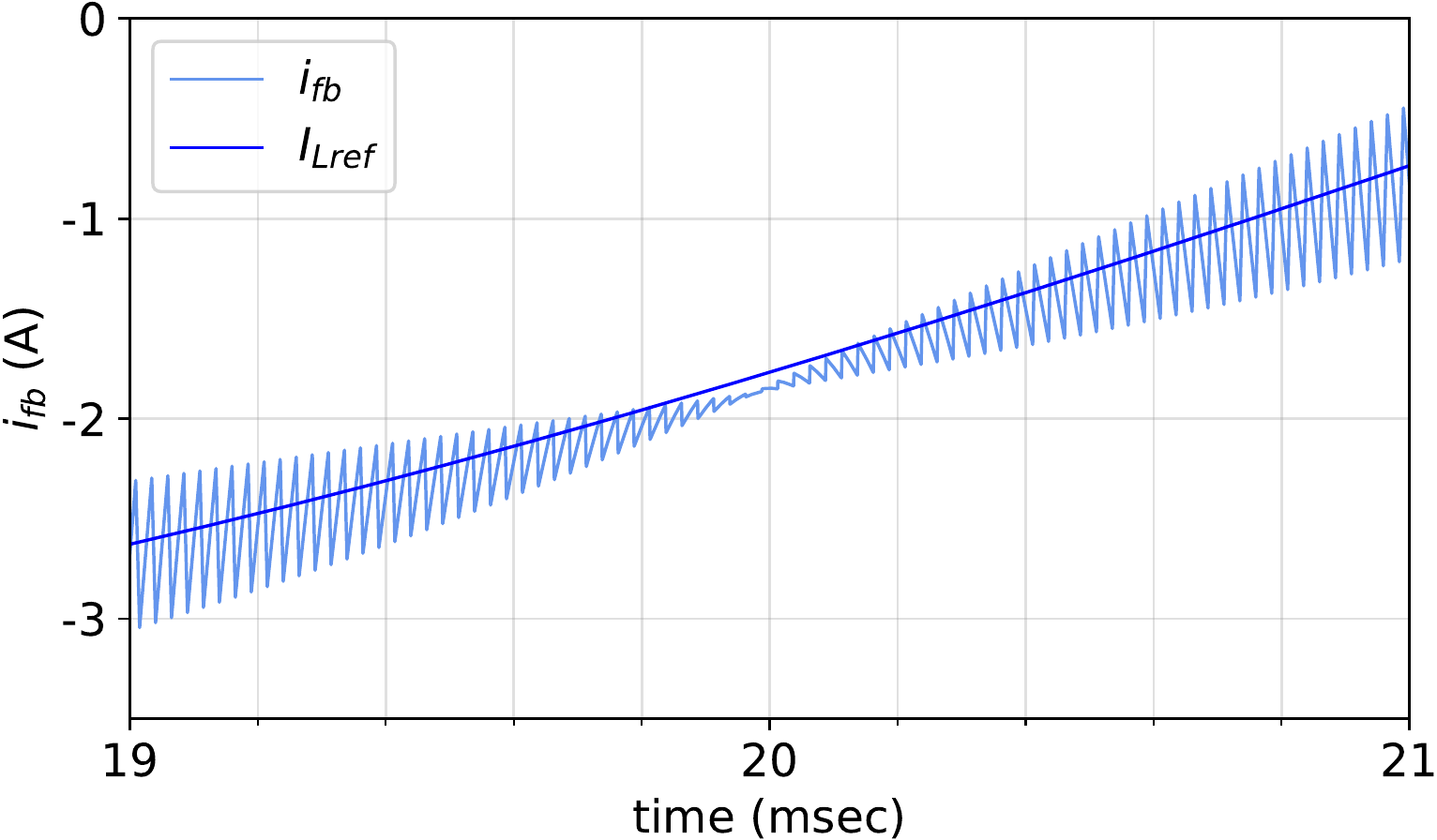}}
\vspace*{-0.7cm}
\caption{ELEX-RKF results for the current-controlled two-level voltage source
converter of Fig.~\ref{fig_cloop1_1} on an expanded scale.}
\label{fig_cloop1_3}
\end{figure}

In the interest of resolving the gate signal waveforms accurately,
the simulator time steps are limited by placing additional time points
before and after the expected cross-over points of the comparators,
as described in \cite{patil2021gseim}.
Simulation results obtained for $i_{fb}(t)$ are shown in
Fig.~\ref{fig_cloop1_2}, along with the reference current
$I_{Lref}(t)$.
The two currents are shown on an expanded scale in
Fig.~\ref{fig_cloop1_3}.

\begin{figure}[!ht]
\centering
{\includegraphics[width=0.49\textwidth]{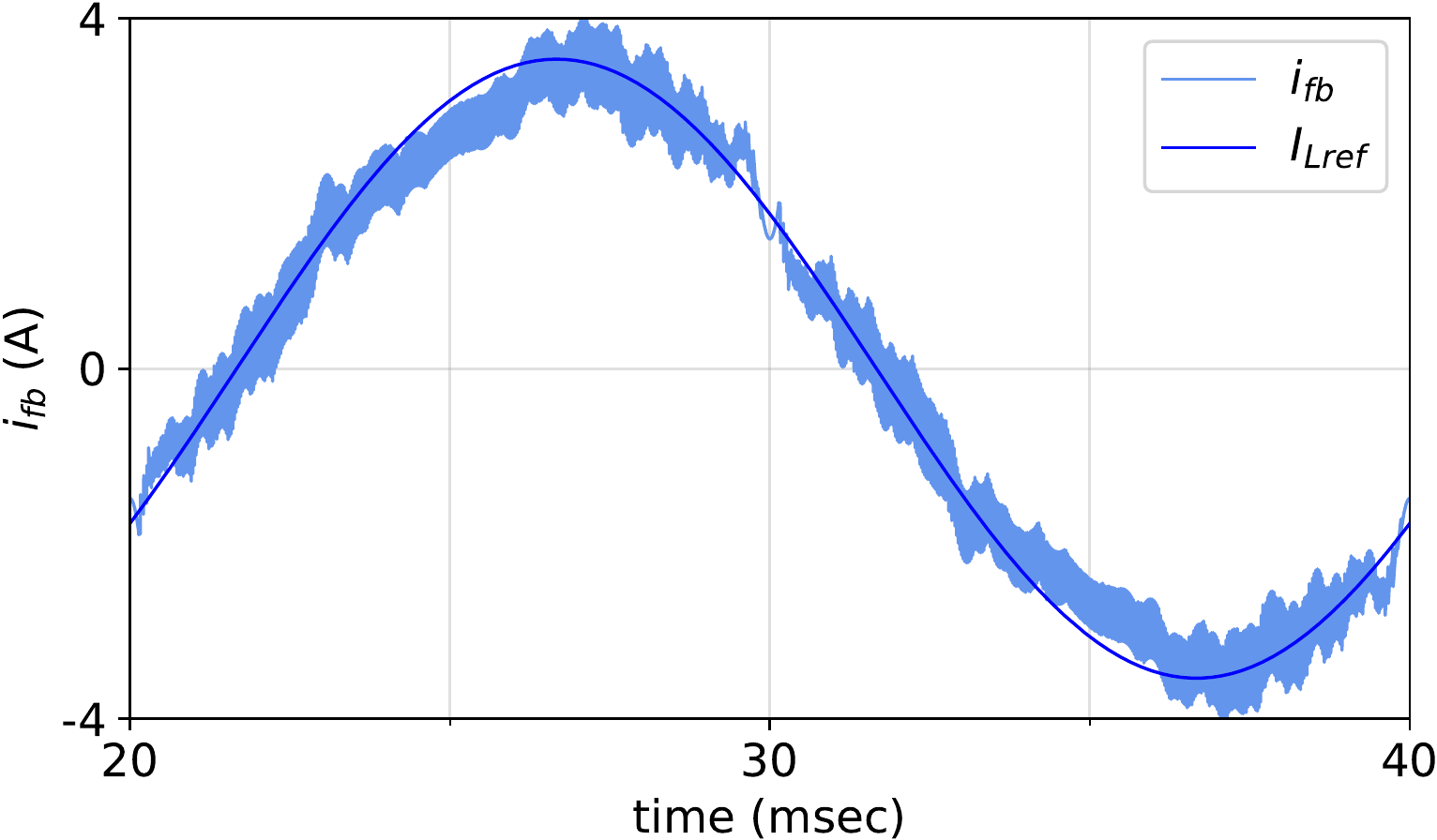}}
\vspace*{-0.7cm}
\caption{ELEX-RKF results for the current-controlled two-level voltage source
converter of Fig.~\ref{fig_cloop1_1}, when comparator
cross-over points are not treated accurately.}
\label{fig_cloop1_4}
\end{figure}

Simulation results obtained for $i_{fb}(t)$,
when the comparator cross-over time points are not treated accurately
(with all other simulation parameters the same),
are shown in
Fig.~\ref{fig_cloop1_4}.
By comparing
Figs.~\ref{fig_cloop1_2} and
\ref{fig_cloop1_4}, we can clearly see that
simulation accuracy is significantly affected by correct treatment
of comparator cross-over.

\subsection{Voltage-controlled buck DC-DC converter}
\label{sec_cloop_buck}
\begin{figure*}[!ht]
\centering
\scalebox{0.9}{\includegraphics{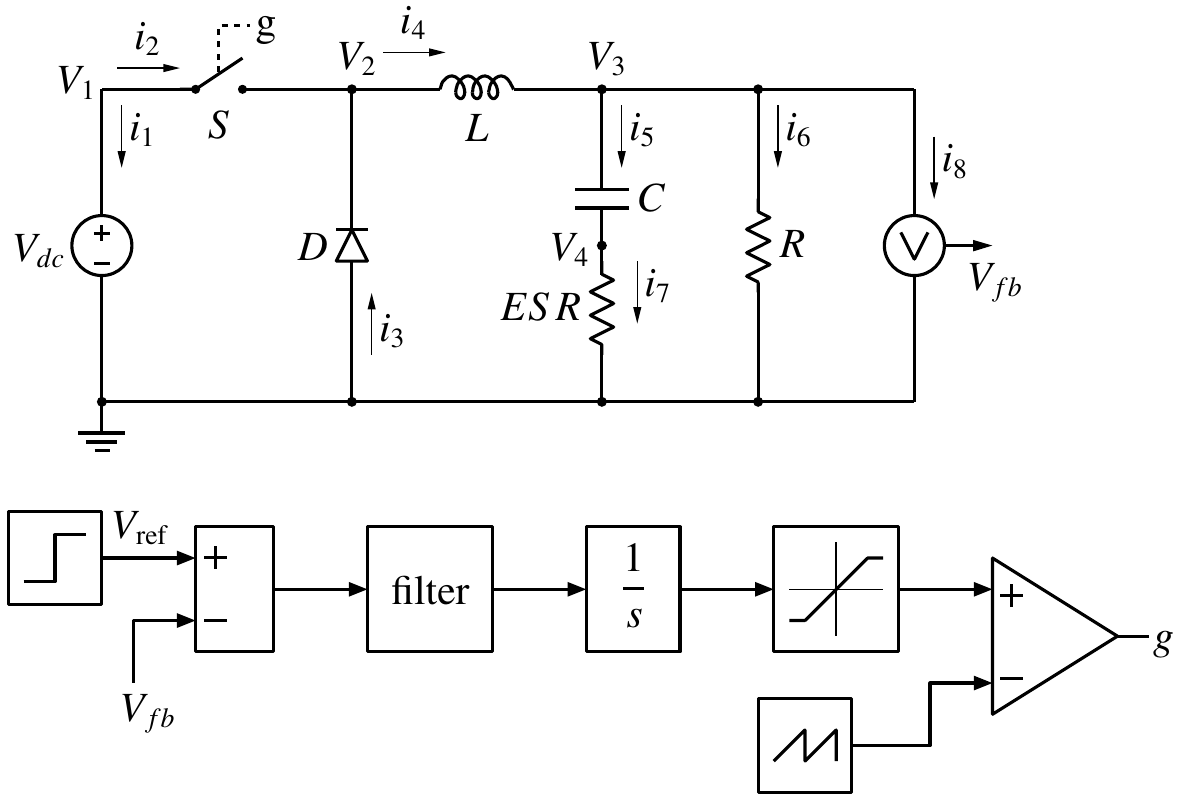}}
\caption{Buck converter circuit and control block. The circuit
parameters are
$V_{dc} \,$=$\, 25$\,V,
$C \,$=$\, 500\,\mu$F,
$L \,$=$\, 24\,\mu$H,
$R \,$=$\, 4\,\Omega$,
$ESR \,$=$\, 0.08\,\Omega$.
The sawtooth waveform has minimum and maximum values of
0 and 1, respectively, and frequency of 400\,kHz.}
\label{fig_cloop2_1}
\end{figure*}

\begin{figure*}[!ht]
\centering
\scalebox{0.9}{\includegraphics{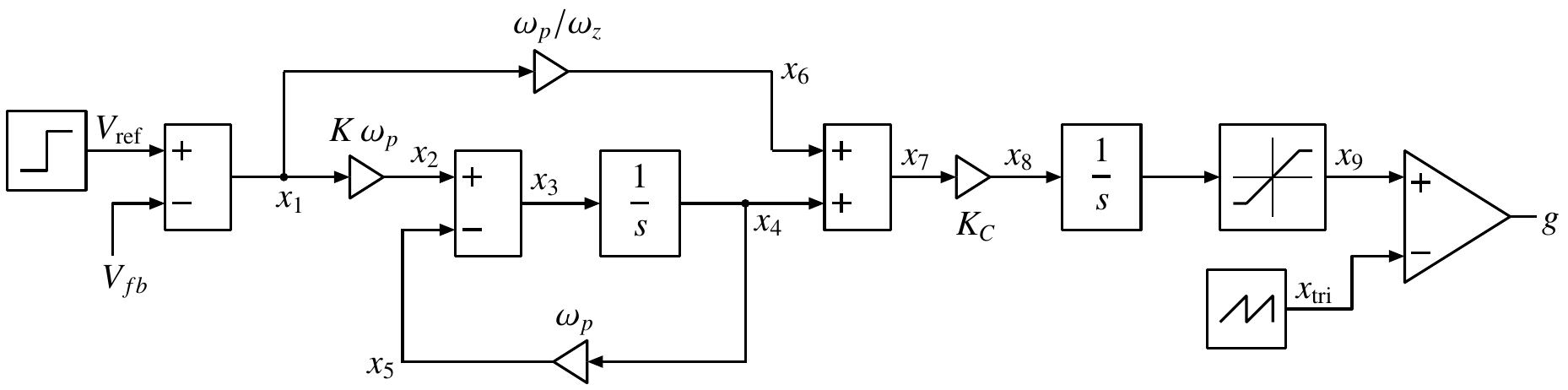}}
\caption{ELEX-RKF implementation of the control block
of the voltage controlled buck converter of
Fig.~\ref{fig_cloop2_1}.}
\label{fig_cloop2_2}
\end{figure*}

Fig.~\ref{fig_cloop2_1} shows a
voltage-controlled buck DC-DC converter, with the voltage reference
$V_{\mathrm{ref}}$ changing from 12\, to 15\,V at $t_{\mathrm{step}} \,$=$\, 10$\,msec.
The transfer function of the filter is given by
\begin{equation}
H(s) = K_C \, \displaystyle\frac{1 + s/\omega _z}{1 + s/\omega _p},
\label{eq_cloop2_1}
\end{equation}
where
$K_C \,$=$\, 4.551 \times 10^3$,
$\omega _z \,$=$\, 6.492\times 10^3$\,rad/s, and
$\omega _p \,$=$\, 6.081\times 10^5$\,rad/s.
For simulation of the circuit using the ELEX-RKF scheme, we rewrite
$H(s)$ as
\begin{equation}
H(s) = K_C \, \left[(\omega _p/\omega _z) + \displaystyle\frac{K}{1 + s/\omega _p}\right],
\label{eq_cloop2_2}
\end{equation}
where
$K \,$=$\, (1-\omega _p/\omega _z)$.
The control block can now be implemented as shown in
Fig.~\ref{fig_cloop2_2}.

As in the previous example, at each stage of the RKF method, we solve
the electrical equations
(${\bf{A}}\,{\bf{x}} \,$=$\, {\bf{b}}$),
followed by updating the control block variables, and then proceed
to the next RKF stage. The control block variables are treated by
first upgrading the two integrator outputs. For example, in RKF stage 2,
we would have
\begin{equation}
x_4^{(2)} = x_{4,n} + h\,\beta _{2,1}\,x_{3,n},
\label{eq_cloop2_3}
\end{equation}
\begin{equation}
x_9^{(2)} = x_{9,n} + h\,\beta _{2,1}\,x_{8,n},
\label{eq_cloop2_4}
\end{equation}
and limit
$x_9^{(2)}$
as $0 \leq x_9^{(2)} \leq 1$.
Next, we update the
$V_{\mathrm{ref}}^{(2)}$ and
$x_{\mathrm{tri}}^{(2)}$ (sawtooth)
source values at
$(t_n + \alpha _2 h)$, and then compute the remaining
control variables as,
\begin{equation}
x_1^{(2)} = V_{\mathrm{ref}}^{(2)} - V_{fb}^{(2)},
\end{equation}
\begin{equation}
x_5^{(2)} = \omega _p\,x_4^{(2)},
\end{equation}
\begin{equation}
\begin{array}{cl}
g^{(2)} = 1 &{\textrm{if}}~x_9^{(2)} > x_{\mathrm{tri}}, \\
\T g^{(2)} = 0 &{\textrm{otherwise}},
\end{array}
\end{equation}
\begin{equation}
x_2^{(2)} = K\,\omega _p x_1^{(2)},
\end{equation}
\begin{equation}
x_6^{(2)} = \displaystyle\frac{\omega _p}{\omega _z}\,x_1^{(2)}
\end{equation}
\begin{equation}
x_3^{(2)} = x_2^{(2)} - x_5^{(2)},
\end{equation}
\begin{equation}
x_7^{(2)} = x_6^{(2)} + x_4^{(2)},
\end{equation}
\begin{equation}
x_8^{(2)} = K_C\,x_7^{(2)}.
\end{equation}

\begin{figure}[!ht]
\centering
{\includegraphics[width=0.49\textwidth]{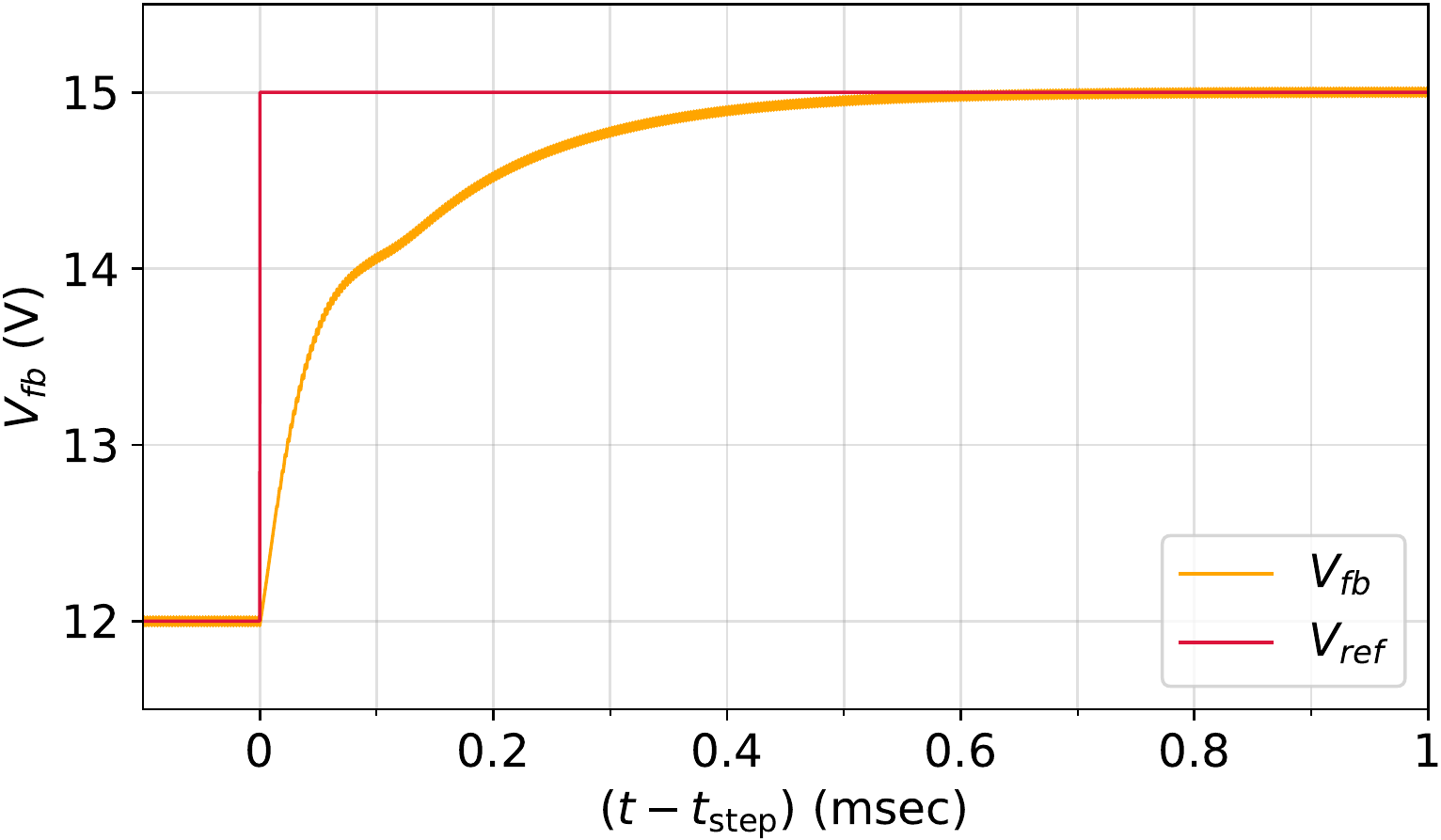}}
\vspace*{-0.7cm}
\caption{Plot of $V_{fb}(t)$ as obtained with the ELEX-RKF scheme for
the voltage controlled buck converter of Fig.~\ref{fig_cloop2_1}.}
\label{fig_cloop2_3}
\end{figure}

\begin{figure}[!ht]
\centering
{\includegraphics[width=0.49\textwidth]{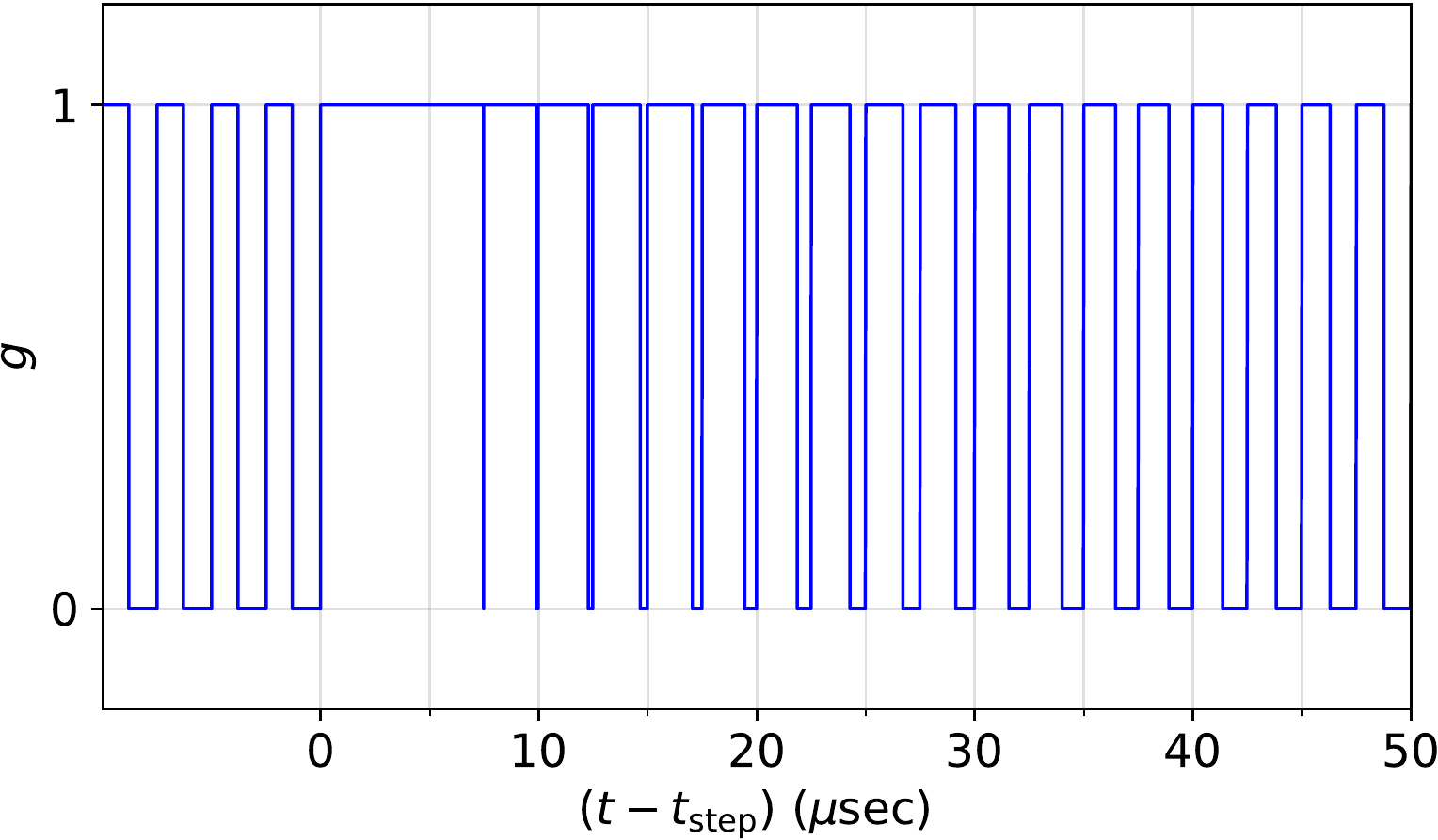}}
\vspace*{-0.7cm}
\caption{Plot of gate signal as obtained with the ELEX-RKF scheme for
the voltage controlled buck converter of Fig.~\ref{fig_cloop2_1}.}
\label{fig_cloop2_4}
\end{figure}

Fig.~\ref{fig_cloop2_3} shows $V_{\mathrm{ref}}(t)$
and $V_{fb}(t)$, and
Fig.~\ref{fig_cloop2_4} shows the gate pulses produced by the
controller when
$V_{\mathrm{ref}}$
changes from 12\,V to 15\,V at $t \,$=$\, t_{\mathrm{step}}$.

\section{Conclusions and future work}
\label{sec_conclusions}
In summary, we have presented a methodology called ELEX-RKF for using
explicit methods~-- in particular the RKF method~-- for simulation of
power electronic circuits, considering the switches to be ideal. The
reasons for higher simulation speed of the ELEX-RKF scheme as compared
to implicit schemes are explained. Several challenges which are encountered
in implementing the ELEX-RKF scheme are pointed out. With the help of suitable
examples, these issues are illustrated, and techniques are descried for
resolving the same. Simulation results obtained using the ELEX-RKF scheme
are presented for a few representative power electronic circuits.

The following future work is planned.
\begin{list}{\arabic{cntr1}.}{\usecounter{cntr1}}
 \item
  The ELEX-RKF scheme will be implemented in the open-source simulation
  package GSEIM\,\cite{gseimarxiv2},\,\cite{gseimgithub}.
  At the time of writing, GSEIM allows only implicit methods for electrical
  circuits, together with MNA for circuit equation formulation.
  Incorporation of the ELEX-RKF option will make GSEIM more versatile,
  and particularly useful for circuits which take a
  substantial amount of time with implicit methods.
 \item
  PLECS\,\cite{plecs} uses the Dormand-Prince (DP) 4-5 pair for solving the
  ODEs. Implementation of the DP method is conceptually very similar to the
  RKF method. It is envisaged that GSEIM will allow both ELEX-RKF and
  ELEX-DP options to the user.
 \item
  In this work, we have not compared simulation times of the ELEX-RKF scheme
  with implicit methods. In future, we plan to use GSEIM to make such a comparison for
  typical power electronic circuits, thus enabling the user
  to make an informed choice about the solution method.
\end{list}

\bibliographystyle{IEEEtran}
\bibliography{ref3}

\end{document}